\documentclass[fleqn,usenatbib]{mnras}

\usepackage{newtxtext,newtxmath}

\usepackage[T1]{fontenc}

\DeclareRobustCommand{\VAN}[3]{#2}
\let\VANthebibliography\thebibliography
\def\thebibliography{\DeclareRobustCommand{\VAN}[3]{##3}\VANthebibliography}

\usepackage{graphicx}
\usepackage{amsmath}	

\title[Emission-line class vs radio properties in RLAGN]{The nature of HERGs and LERGs in LoTSS DR2 $-$ a morphological perspective}

\author[J. Chilufya et al.]{
J. Chilufya,$^{1}$
\thanks{E-mail: \href{mailto:j.chilufya@herts.ac.uk}{j.chilufya@herts.ac.uk}}
M. J. Hardcastle,$^{1}$
J. C. S. Pierce,$^{1}$
A. B. Drake,$^{1}$
R. D. Baldi,$^{2}$
H. J. A. R\"ottgering$^{3}$
\newauthor and D. J. B. Smith$^{1}$\\
$^{1}$Centre for Astrophysics Research, Department of Physics, Astronomy and Mathematics, University of Hertfordshire, Hatfield AL10 9AB, UK\\
$^{2}$INAF-Istituto di Radioastronomia, Via Piero Gobetti 101, 40129, Bologna, Italy\\
$^{3}$Leiden Observatory, Leiden University, PO Box 9513, NL-2300 RA Leiden, The Netherlands\\
}

\pubyear{2023}

\begin{document}
\label{firstpage}
\pagerange{\pageref{firstpage}--\pageref{lastpage}}
\maketitle

\begin{abstract}
We present the largest visually selected sample of extended ($>$60 arcsec) radio-loud active galactic nuclei (RLAGN) to date, based on the LOw-Frequency Array Two-Metre Sky Survey second data release (LoTSS DR2). From the broader LoTSS DR2 dataset with spectroscopic classifications, we construct a subsample of 2828 RLAGN with radio luminosities greater than $10^{23}~\mathrm{W~Hz^{-1}}$ at $z<0.57$. These RLAGN are further classified by optical emission-line properties into high-excitation and low-excitation radio galaxies, enabling a detailed emission-line analysis. Our subsample is also morphologically classified into Fanaroff \& Riley centre- and edge-brightened (FRI/FRII) sources, wide- and narrow-angle tail (WAT and NAT) sources, head-tail (HT) sources, and relaxed double (RD) sources. For these classifications, we utilize data from the Very Large Array Sky Survey (VLASS) to assist with the classification, taking advantage of its 2.5 arcsec resolution which is sensitive to structures below 30 arcsec. This resolution allows us to identify compact cores and hotspots, facilitating the identification of remnant and restarted RLAGN candidates. We investigate the relationship between emission-line and radio properties in RLAGN, analyzing mid-infrared data, host galaxy mass, and core prominence. These analyses uncover the complex relationship between these factors and the underlying accretion mechanisms. Our findings emphasize that no single property can fully constrain the accretion mode in RLAGN, highlighting the necessity of multi-dimensional approaches to reveal the processes driving RLAGN behaviour.
\end{abstract}

\begin{keywords}
galaxies: active -- galaxies: jets -- radio continuum: galaxies.
\end{keywords}

\section{Introduction}\label{sec:into} 
Radio observations favour the identification of a subpopulation of active galactic nuclei (AGNs) primarily dominated by jets, so-called radio-loud AGNs (RLAGN). RLAGN exhibit distinctive features such as cores, hotspots, jets, and lobes, which allow their classification into various morphological types. These include (but are not limited to) the Fanaroff-Riley type I and II (FRI/FRII) objects \citep{1974MNRAS.167P..31F}, wide-angle tails \citep[WATs;][]{1976ApJ...205L...1O}, narrow-angle tails \citep[NATs;][]{1976ApJ...203L.107R,1980ARA&A..18..165M}, head-tails \citep[HTs;][]{1968MNRAS.138....1R}, and relaxed doubles \citep[RDs;][]{1989MNRAS.238..357O,Leah93}. These morphological differences provide both theorists and observers with a distinctive perspective on how radio jets are influenced by their surroundings \citep[e.g.][]{2018MNRAS.476.1614C,2019A&A...622A..10C}. 

Many FRIs are found in groups of galaxies where the interaction with the intracluster medium (ICM) causes their jets to decelerate and spread out \citep[e.g.][]{1994ApJ...422..542B,2008MNRAS.386.1709C}, resulting in center-brightened structures \citep{1974MNRAS.167P..31F}. In contrast, FRIIs are often located in less dense environments \citep[e.g.][]{2000MNRAS.314..359H}, allowing their jets to remain collimated and form bright hotspots at the edges, creating edge-brightened lobes \citep{1974MNRAS.167P..31F}. WATs and NATs are influenced by the motion of their host galaxies through the cluster medium \citep[e.g.][]{1972Natur.237..269M,1979Natur.279..770B}, with WATs having wide, gently curved tails, and NATs showing narrow, sharply bent tails \citep[e.g.][]{2019MNRAS.488.2701M,2020MNRAS.496..676B} due to strong ram pressure \citep[e.g.][]{1994ApJ...436...67V}. HTs are characterized by their unique head-tail structures caused by the combined effects of galaxy motion and interaction with the surrounding medium \citep[e.g.][]{2021MNRAS.508.5326M,2024MNRAS.528..141L} and are sometimes thought of as unresolved NATs \citep[e.g.][]{2017A&A...608A..58T}. RDs, on the other hand, display more relaxed and symmetric structures, suggesting less interaction with dense environments \citep{1989MNRAS.238..357O}. 

Understanding the underlying accretion processes behind these different morphological types of RLAGN requires optical -- and sometimes mid-infrared -- observations. These observations categorize RLAGN into two types, regardless of radio morphology, based on their optical emission-line properties \citep{1979MNRAS.188..111H}: high-excitation radio galaxies and low-excitation radio galaxies (hereafter HERGs and LERGs respectively). The key differences between HERGs and LERGs, in the local Universe in particular, can be viewed in terms of their fuelling mechanism, accretion rate, ionization, jets and lobes, and their host galaxies \citep[e.g.][]{2020NewAR..8801539H}.

HERGs are primarily driven by accretion processes that are radiatively efficient (RE), typically fueled by cold gas reservoirs \citep[e.g.][]{2007MNRAS.376.1849H}. This cold gas can arise from sources such as merger interactions \citep[e.g.][]{2012MNRAS.419..687R,2019MNRAS.487.5490P,2022MNRAS.510.1163P,2023MNRAS.522.1736P} or cooling flows from the intergalactic medium \citep[IGM; e.g.][]{2018NatAs...2..273H}. Recent studies, however, suggest that while cold gas accretion typically results in high accretion rates, this is not an absolute requirement for RE AGNs \citep[e.g.][]{2020NewAR..8801539H}. In some cases, cooling instabilities in the hot gas can also drive RE accretion. These high accretion rates lead to the formation of a geometrically thin, optically thick accretion disk \citep[e.g.][]{1973A&A....24..337S}. The associated strong radiation ionises the surrounding gas, which in turn emits prominent high-excitation emission lines such as [O{\scriptsize{III}}] and [N{\scriptsize{II}}] \citep[e.g.][]{2009A&A...495.1033B,2010A&A...509A...6B}, consistent with the standard AGN model \citep{1993ARA&A..31..473A}. In this model, the central engine consists of a supermassive black hole surrounded by the dusty torus and accretion disk, with differences in AGN observations primarily explained by orientation effects. Depending on the viewing angle, the dusty torus may obscure or reveal the narrow- and broad-line regions, resulting in the observed diversity in AGN classifications.

LERGs, on the other hand, are primarily associated with radiatively inefficient (RI) accretion processes, commonly fueled by hot gas from the halo of the galaxy or the ICM \citep[e.g.][]{2006MNRAS.372...21A,2012ARA&A..50..455F}. However, recent studies also indicate that RI AGNs can be sustained by low-level accretion from cold gas reservoirs \citep[e.g.][]{2018NatAs...2..273H,2020NewAR..8801539H}. The fueling is often mediated by relatively stable accretion dynamics, such as those described by Advection-Dominated Accretion Flows (ADAFs) or Radiatively Inefficient Accretion Flows (RIAFs) \citep[e.g.][]{1994ApJ...428L..13N,1995ApJ...444..231N}. The lower accretion rates result in minimal ionising radiation from the central engine, leading to weak low-excitation emission lines \citep{2006MNRAS.370.1893H,2012MNRAS.421.1569B}. Despite these differences, both accretion modes highlight the importance of Eddington-scaled accretion rates in determining radiative efficiency, rather than exclusively the temperature or phase of the accreted gas.

In terms of their host environments, HERGs and LERGs have a clear tendency to be in different environments \citep[e.g.][]{2009ApJ...696...24S,2012MNRAS.421.1569B}, particularly at lower redshifts. HERGs often favour younger, more actively star-forming galaxies (SFGs), sometimes where galaxy mergers or interactions are common. LERGs, however, typically live in older, more massive elliptical galaxies with little to no ongoing star formation (SF), although this is not the case at higher redshifts \citep[$z>1$; e.g.][]{2022MNRAS.513.3742K,2025MNRAS.536..554K}. These galaxies are often located in dense environments such as galaxy clusters, where the hot gas reservoir is more readily available.
  
Several studies have attempted to associate different morphological types of RLAGN with the two modes of excitation \citep[e.g.][]{1997MNRAS.286..241J,1999MNRAS.304..160J}. For instance, the majority of FRIs are associated with LERGs, found in dense environments where the lower energy output from the AGN leads to broader, less collimated jets that interact heavily with the ICM. Conversely, the majority of FRIIs are often linked with HERGs, where higher energy outputs maintain well-collimated jets that extend to form bright hotspots. However, this association is not entirely the case. Even in the extensively studied 3CRR sample \citep{1983MNRAS.204..151L}, which has long served as a foundation for understanding RLAGN, there have always been clear examples of FRIIs with a LERG spectrum. 

While WATs, NATs, and HTs are predominately classified as LERGs \citep[e.g.][]{2019A&A...626A...8M}, these are often interpreted as bent FRIs, with their morphology shaped more by the host galaxy’s motion through the ICM than excitation properties. On the other hand, objects classified as RDs are generally expected to be LERGs, as their nuclear activity has diminished, resulting in weak or no optical excitation. The resemblance of RD lobes to those of FRIIs -- albeit without the characteristic features such as compact cores, jets, and hotspots -- has led to suggestions that RDs represent a late evolutionary stage of radio galaxies, entering the so-called remnant phase. \citep[e.g.][]{2016MNRAS.462.1910H,2018MNRAS.475.4557M,2018MNRAS.475.2768H}. 

Potentially following the remnant phase is the restarted phase of AGN activity \citep[e.g.,][]{2019A&A...622A..13M, 2020A&A...638A..34J}, marked by the reactivation of the central engine. This phase is distinguished by new nuclear activity that propagates through the remnant lobes, often creating structures indicative of renewed jet activity. These restarted jets may manifest as compact, double-lobed features embedded within or surrounded by the older, more extended remnant lobes 
\citep{2000MNRAS.315..371S,2024arXiv240813607D}. In some cases, the re-ignited AGN activity triggers ionization in the surrounding gas, which can result in restarted objects exhibiting a HERG-like spectrum. This phase not only redefines the morphology of the radio galaxy but also provides insights into the episodic nature of AGN activity \citep[e.g.][and references therein]{2023Galax..11...74M}, where periods of quiescence (on Myrs) are followed by renewed nuclear activity. Therefore, understanding all these associations helps piece together the environmental and internal mechanisms that shape the observed radio morphologies and the excitation states of these AGNs.

While attempts to establish the relationship between radio power and line emission properties have been made \citep[e.g.][]{2004ApJ...613..109H,2006MNRAS.370.1893H,2007MNRAS.376.1849H,2009MNRAS.396.1929H,2014MNRAS.440..269M}, a clear-cut relationship between radio properties and emission line properties is yet to be seen \citep[e.g.][]{2013MNRAS.430.3086G,2022MNRAS.511.3250M}. This ambiguity suggests that the processes governing the radio emission and the line excitation in AGNs might be more complex than initially thought \citep[e.g.][]{1996AJ....112....9L,2001A&A...379L...1G}. Several factors contribute to this complexity, including environmental conditions, host galaxy properties, accretion modes, orientation effects, feedback processes, and constraints in RLAGN samples due to survey sensitivity limitations \citep{2016A&ARv..24...13P}.

Thanks to modern state of the art radio telescopes such as the LOw-Frequency Array \citep[LOFAR;][]{2013A&A...556A...2V} and the NRAO {\it Karl G Jansky} Very Large Array \citep[VLA;][]{2011ApJ...739L...1P}, we are able to probe deeper into these complexities with unprecedented sensitivity and resolution. Deep sky surveys such as the LOFAR Two-metre Sky Survey \citep[LoTSS;][]{2017A&A...598A.104S,2019A&A...622A...1S,2022A&A...659A...1S} and the VLA Sky Survey \citep[VLASS;][]{2020PASP..132c5001L} allow us to observe RLAGN across a broad range of frequencies and scales. These capabilities enable detailed studies of the faintest radio emission \citep[e.g.][]{2019A&A...622A..17S,2019A&A...622A..12H}, the life cycles of radio jets \citep[e.g.][]{2017A&A...606A..98B,2020A&A...638A..34J}, and the environments of RLAGN \citep[e.g.][]{2019A&A...622A..10C}.

Motivated by the need to investigate the relationship between radio properties and emission-line properties, we have, for the first time, selected a large sample of resolved RLAGN with emission-line classifications based on the LoTSS second data release \citep[DR2;][]{2022A&A...659A...1S}. We classified these RLAGN into various morphological types and separated them into HERG and LERG categories. The aims of this paper are threefold: we will (1) visually construct a large sample of extended RLAGN from LoTSS DR2; (2) categorize the RLAGN into HERG/LERG based on their line properties; and (3) explore the relationship between emission-line properties and radio properties.

The structure of this paper is as follows: Section \ref{sec:data} describes the data used, while Section \ref{sec:samp} outlines the sample selection and classification of RLAGN into various morphological types, including FRI, FRII, WAT, NAT, HT, and RD sources. This section also explains the separation of these objects into the two accretion modes, HERG and LERG. Section \ref{sec:results} presents the results, followed by a discussion in Section \ref{sec:discusion}, and conclusions in Section \ref{sec:concl}. 

Where we quote physical parameters, these have been computed assuming a $\Lambda$CDM cosmological model with the following values: $H_0 = 70~\rm{km~s^{-1}~Mpc^{-1}}$, $\Omega_{\rm{m0}} = 0.3$, and $\Omega_{\Lambda0} = 0.7$. The spectral index is assumed in the sense $S_{\nu}\propto \nu^{-\alpha}$.

\section{DATA}\label{sec:data}
In this work, we aim to build a sample of extended RLAGN with different morphological classifications: FRI, FRII, WAT, NAT, HT, and RD sources using LoTSS DR2 data.

DR2 \citep{2022A&A...659A...1S} covers two distinct regions: the first is centered at $\rm{12^h45^m}$ right ascension and $+44^{\circ}30^{\prime}$ declination, while the second is centered at $\rm{1^h00^m}$ right ascension and $+28^{\circ}00^{\prime}$ declination, spanning 4178 and 1457 degree$^2$, respectively. The median sensitivity of DR2 is $\sim$83 $\mu$Jy per beam, achieved at a resolution of 6 arcsec (144 MHz). We refer the reader to \citet{2022A&A...659A...1S} for details of the techniques employed in constructing a sample comprising 4 396 228 extragalactic radio sources which include both AGNs and SFGs.

The procedure of assigning optical counterparts to the DR2 catalogue is presented by \citet{2023A&A...678A.151H}, which follows the procedures similar to those that were applied to LoTSS DR1 \citep{2019A&A...622A...2W,2021A&A...648A...4D}. The resulting sample consists of $\sim$4.2 million radio sources detected with $ \geq5\sigma$ significance, with 85 per cent of them having optical counterparts \citep{2023A&A...678A.151H}. Additionally, 58 per cent of these sources have reliable redshift estimates, obtained through either spectroscopic or photometric methods \citep{2021A&A...648A...4D}.

To address the objectives of this paper as outlined in Sec. \ref{sec:into}, we utilized a sample consisting of 152 355 radio sources \citep[][hereafter D24]{2024MNRAS.534.1107D} created by cross-matching the DR2 sources with the Portsmouth Sloan Digital Sky Survey catalogue \citep{2013MNRAS.431.1383T}. These sources cover the redshift range between $0.0<z<0.57$. D24 compiled a sample of DR2 sources with emission-line measurements and classified them into distinct radio source categories, including SFGs, radio-quiet AGNs (RQAGN), and HERGs or LERGs. Detailed procedures utilized in this classification process are presented in D24; however, a brief description is provided here, as these methodologies are essential for constructing a subsample to address our objectives at hand.

To integrate spectroscopic information with the low-frequency radio data, the positions of optical counterparts identified for the radio sources by \cite{2023A&A...678A.151H} were matched with the Portsmouth spectroscopic catalogue using a nearest-neighbour algorithm with a maximum search radius of one arcsec. This process yielded a subset of 208 816 LOFAR sources with SDSS spectroscopic information.

A series of cuts and filters were applied to refine the initial matched catalogue. Redshift cuts removed sources at redshifts where H$\alpha$ is not visible to the SDSS spectrographs: that is, at $z=0.385$ for SDSS DR8 and $z=0.570$ for BOSS. Additionally, sources with zero, or missing H$\alpha$ fluxes and associated errors, as well as with nonphysical Balmer line luminosities, were excluded by considering only those with $4<\log_{10}(L_{\mathrm{H\alpha}}/L_{\odot})<13$. To limit the influence of sources whose radio flux is significantly larger than what is captured in the SDSS aperture, sources with $r_{50}>15$ arcsec, the radius within which 50 per cent of the total flux is contained, were removed. For the final catalogue, sources missing any of the important BPT diagnostic emission lines such as H$\alpha$, H$\beta$, [O{\scriptsize{III}}], [N{\scriptsize{II}}] \citep{1981PASP...93....5B,2009A&A...495.1033B,2010A&A...509A...6B,2012MNRAS.421.1569B} were excluded, resulting in a subset of 124 023 sources with complete spectral measurements.

The classification involved identifying radio excess sources by comparing the 144 MHz luminosity with the H${\alpha}$ luminosity. For sources with complete diagnostic emission lines, the BPT diagram was used to distinguish between SFGs and AGNs, further classifying them into radio-quiet AGNs (RQAGN), HERGs or LERGs. To quantify the reliability of each classification, Monte-Carlo simulations were performed, generating 1000 realizations of the matched catalogue. The results were combined into an overall probabilistic set of classifications, accounting for uncertainties and providing a probabilistic reliability for each source's classification. Applying a 90 per cent ($p>0.9$) reliability threshold, D24 compiled a final classification using this method, resulting in the identification of 38 728 SFGs, 18 726 RQAGN, 38 588 radio-excess AGNs, 362 HERGs, and 12 648 LERGs.

For sources lacking the four diagnostic emission lines necessary for methods like the BPT diagnostic, an alternative approach was described by D24 that allows the selection of additional HERGs. This method involved using the specific [O{\scriptsize{III}}] luminosity as a proxy for the radiative activity of AGNs \citep{1994ASPC...54..201L}. This method was based on the understanding that the HERG/LERG dichotomy reflects a transition between radiatively efficient and inefficient AGN activity occurring at a few per cent of the Eddington rate \citep[e.g.][]{2012MNRAS.421.1569B}. By examining radio-excess objects, the luminosity in the [O{\scriptsize{III}}] line was used as an indicator, with standard linear corrections applied to approximate bolometric luminosities \citep{2004ApJ...613..109H}. Although black hole mass ($M_{\mathrm{BH}}$) estimates are not available, galaxy mass/$M_{\mathrm{BH}}$ mass relations were utilized as proxies. The galaxy mass estimates given by \cite{2023A&A...678A.151H}, derived from the DESI Legacy Survey and WISE photometry, were preferred over those based on SDSS photometry.

A histogram of the specific [O{\scriptsize{III}}] luminosity for radio-excess objects, applying $p>0.9$, was created, broken down into HERG and LERG classes (we return to this point in Sec. \ref{sec:samp}). The results indicated that HERGs typically have higher specific [O{\scriptsize{III}}] luminosity, aligning with previous findings, although there was substantial overlap. This method validated the classification and helped identify objects lacking the necessary lines for a full BPT classification.

We apply this method, combined with the reliability threshold approach, to classify our subsample into HERG and LERG to address the scientific objectives of this paper.

\section{RLAGN Classifications} \label{sec:samp}
\subsection{Radio morphology classifications}\label{sec:radio_morp}
\begin{table}
\centering
    \caption{Summary of results based on visual morphological classifications.}
    \begin{tabular}{c|c|c}
    \hline \hline
     Morphology & \# of objects with 3 votes & \# of objects with 2 votes  \\
     \hline
     FRI  & 202 & 634\\
     FRII & 563 & 408\\
     WAT & 122 & 359\\
     NAT & 51 &  110\\
     HT & 136 & 154\\
     RD & 26 & 128\\
     \hline
     Total & 1100 & 1793\\
     \hline\hline
    \end{tabular}
    \label{tab:sum_morph}
\end{table}

Categories for extended RLAGN (FRI, FRII, WAT, NAT, HT, and RD sources) were systematically classified using visual inspection techniques applied to the radio-excess ($p>0.9$) sample from D24. We first focused on large resolved objects whose components were identified using the Python Blob Detection and Source Finder \citep[PyBDSF;][]{2015ascl.soft02007M} and associated using the process described by \cite{2023A&A...678A.151H}. Requiring total flux densities $>$10 mJy, and physical size estimates $>$60 arcsec ensured that we identified objects suitable for visual classification. From a preliminary D24 sample consisting of 152 355 radio objects with optical cross-matches and good redshift estimates \citep[as reported by][]{2023A&A...678A.151H}, filtering based on size and flux density left us with 4647 candidates identified as RLAGN. Where available, we utilized multiple high-frequency VLASS images \citep{2020PASP..132c5001L,2021ApJS..255...30G} of the same target, stacking them at the radio source positions to produce deeper images. These enhanced images facilitated the robust identification of ongoing RLAGN activity, including compact cores, hotspots, and evidence of restarts. This information will be valuable in later sections. VLASS offers several advantages over other radio surveys such as the Faint Images of the Radio Sky at Twenty-Centimeters \citep[FIRST;][]{1995ApJ...450..559B} and NRAO VLA Sky Survey \citep[NVSS;][]{1998AJ....115.1693C}. For example, VLASS offers a significantly higher resolution (2.5 arcsec) compared to FIRST (5 arcsec) and NVSS (45 arcsec). With a median sensitivity of approximately 70 $\mu$Jy at 2.5 arcsec, VLASS can detect and resolve flat-spectrum cores in LOFAR objects that remain unresolved due to LOFAR's sensitivity to steep-spectrum emission. This enhances the ability to image RLAGN structures in greater detail. 

We then examined DR2 objects with the aid of both LoTSS and VLASS images for distinctive features (cores, hotspots, lobes, jets, wings, plumes, and so on) to determine whether they belong to the FRI, FRII, WAT, NAT, HT, or RD categories. We describe the criteria used to select these objects below, and a representative sample following the classification process is presented in Fig. \ref{fig:images} and \ref{fig:images2}.

\begin{itemize}
    \item [(i)] \textbf{FRI:} The traditional classification of FRIs and FRIIs was based on morphological features \citep{1974MNRAS.167P..31F} which we adopt here. We identified FRI and FRII sources by examining key structural features in the DR2 images. For FRIs, we looked for jet activity that is brightest near the core and gradually fades with distance, often accompanied by diffuse, edge-darkened lobes.

    \item[(ii)] \textbf{FRII:} These sources feature jets that remain collimated and terminate in bright hotspots located at the edges of the radio lobes. For candidate FRIIs, we looked for distinct edge-brightened lobes with prominent terminal hotspots.

\item[(iii)] \textbf{WAT:} We identified WAT candidates based on the curvature and morphology of their radio jets, characterized by wide `C'-shaped, curved tails. We also looked for other obvious characteristics such as hotspots near the core. These usually appear less bright than the central core.

\item[(iv)] \textbf{NAT:} In contrast, NATs were identified by their distinctive, narrow `V'-shaped tail structures, with jets exhibiting pronounced, narrow-angle bending. To distinguish NATs from WATs, we classified only those objects with visibly narrower angles between their tails as NATs. Additionally, we excluded objects with detectable hotspots from this category.  

\item [(v)] \textbf{HT:} We classified objects as HTs if they show signs of having a bright `head' near the galaxy core and a `tail' that extends outward.

\item[(vi)] \textbf{RD:}  We identified RDs by selecting sources with FRII-like double-lobed structures that appear symmetric but lack pronounced hotspots or jets.
\end{itemize}

Table \ref{tab:sum_morph} presents the results of the visual classification performed by three independent voters, detailing the distribution across categories. Of the 4647 objects, 2893 ($\sim$62 per cent) received at least two votes for the same category. The remaining $\sim$38 per cent received only a single vote in any given category, meaning that while all three voters assessed these objects, their classifications differed, preventing consensus. Additionally, 588 objects were rejected entirely due to exhibiting signs of SF\footnote{Large spirals appear in the SF sample of D24. This is likely due to their high SFRs which amplify radio emission. Additionally, the morphological features of large spirals, such as extended disk structures, enhance their detectability in radio surveys. Bright disks could be mistaken for AGN-related emission as well.}, often appearing featureless in the LOFAR cutout images (see Fig. \ref{fig:example_source}). The 1100 objects that received three votes per category thus indicate robust categorization.

\begin{figure}
    \centering
    \includegraphics[width=\columnwidth]{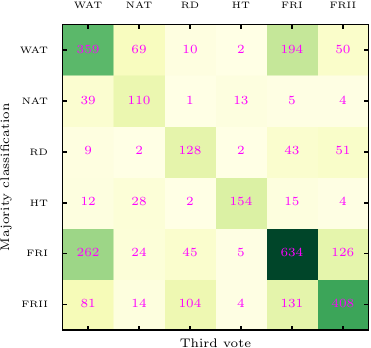}
    \caption{The heatmap shows a strong diagonal dominance, indicating that most objects received the same classification from two voters. The most significant densities are observed along the diagonal, which suggests consistency among the voters, whereas the off-diagonal cells represent cases where objects received mixed classifications, indicating ambiguity or transitional characteristics. For example, some WATs were also voted as FRIs, and some relaxed sources were voted as FRIIs.}
    \label{fig:matrix}
\end{figure}

\begin{figure}
    \centering
    \includegraphics[width=0.49\columnwidth]{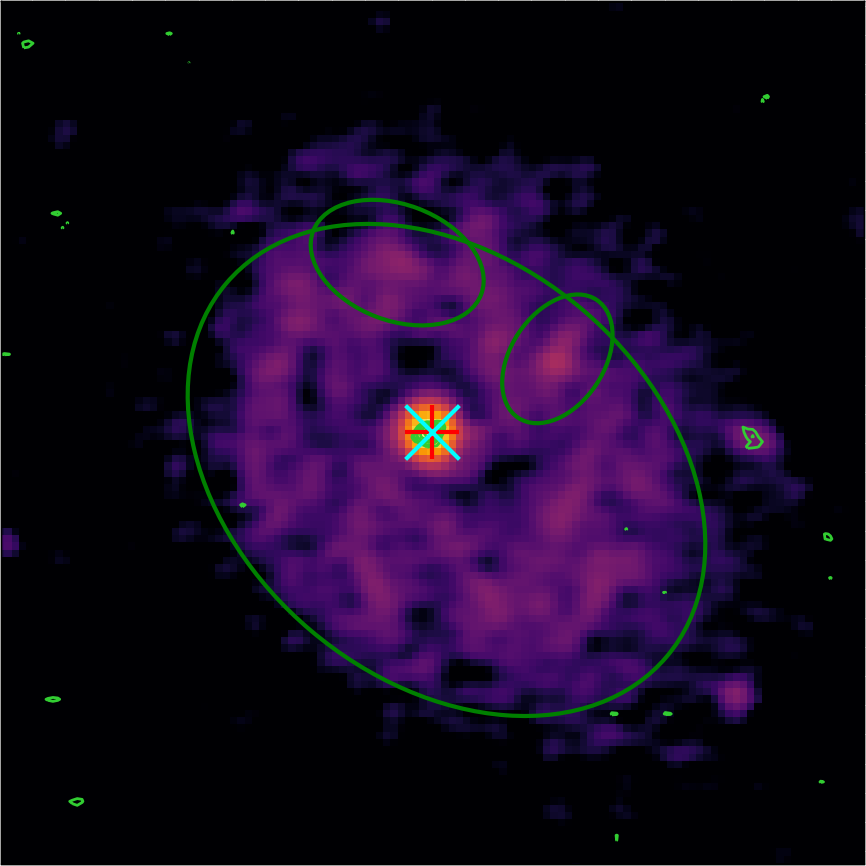}
    \includegraphics[width=0.49\columnwidth]{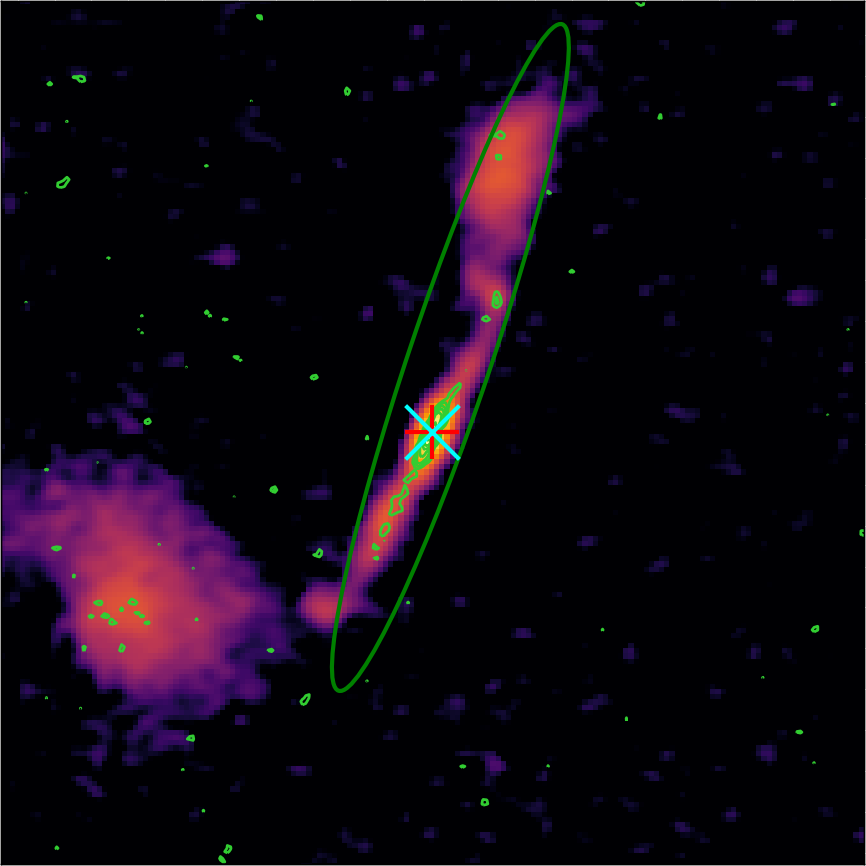}
    \caption{Example images used in the visual classification, showing VLASS contours (green) overlaid on the LOFAR image. The ellipses indicate source component identification by PyBDSF, the cyan cross marks the optical position, and the red plus sign denotes the centroid of the radio source. As noted in the caption of Fig. \ref{fig:matrix}, the right image represents an object that received two votes for the FRI category, while the third vote was assigned to the WAT category. However, the two FRI votes, along with further inspection, were sufficient to classify it as an FRI in our final selection. The image on the left represents objects that were rejected due to SF signatures, as explained in the text.}
    \label{fig:example_source}
\end{figure}

For the objects that received two votes per category, we conducted a further inspection to ensure clarity in their classification. Upon thorough examination and with the aid of Fig. \ref{fig:matrix}, we determined that the sample of objects with two votes was adequate for inclusion. Consequently, we added these objects to our final catalogue alongside those with three votes, resulting in a subsample comprising 2893 RLAGN (see Table \ref{tab:final_samp}).

Fig. \ref{fig:matrix} presents a matrix representation of classification consistency across various RLAGN categories. It shows the distribution of sources for each combination of the majority classification vote and the third vote. Here, the majority classification refers to objects that received two votes, indicating agreement among classifiers on the primary classification. For example, 634 objects received two votes for the FRI category (see an example object shown in Fig. \ref{fig:example_source}), demonstrating a good level of agreement among the classifiers. Similarly, significant numbers of objects received two votes in other categories such as FRII (408), WAT (359), and NAT (110). The figure also shows the distribution of the third vote across different categories, illustrating the level of ambiguity or secondary classification considerations for each RLAGN type.

For instance, we observed that the FRI category had a high level of agreement with 634 objects receiving two votes, with the third vote often leaning towards WAT. The FRII category also showed strong agreement with 408 objects receiving two votes, where the extra vote frequently pointed towards the FRI and relaxed category. WAT and NAT categories also presented a substantial number of majority classifications (359 and 110, respectively), with some overlap in votes with other categories like FRI and HT sources.

This detailed classification breakdown, combined with our rigorous inspection, confirms the robustness of our final sample. We ensured that objects excluded from our final catalogue were indeed challenging to visually classify, thereby validating our exclusion criteria and ensuring the reliability and accuracy of our RLAGN classifications.

\begin{table}
    \centering
    \caption{Morphological break down of the sample following visual classification.}
    \begin{tabular}{c|c}
    \hline \hline
     Morphology & Total \# of objects  \\
     \hline
     FRI  & 836\\
     FRII & 971\\
     WAT & 481\\
     NAT & 161\\
     HT & 290\\
     RD & 154\\
     \hline
     Total & 2893\\
     \hline\hline
    \end{tabular}
    \label{tab:final_samp}
\end{table}

\begin{figure}
    \centering
    \includegraphics[width=\columnwidth]{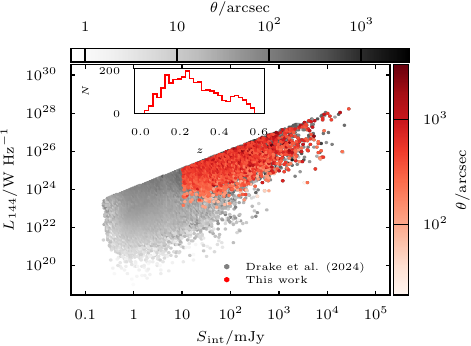}
    \caption{The distribution of our sample is shown in the radio luminosity versus total flux density plane overlaid on the broader D24 sample. The horizontal and vertical colour bars illustrate the angular size of the D24 sample and our sample, respectively.}
    \label{fig:agn_selection}
\end{figure}

Before we introduce the next section, we refined our morphological subsample of 2893 RLAGN by excluding an additional 65 objects with $L_{144}<10^{23}~\rm{W~Hz^{-1}}$. Below these luminosities, the contribution from SF at 144 MHz could be important  for the radio luminosity \citep[e.g.][]{2019A&A...622A..17S,2019A&A...622A..12H}, so this cut was applied to mitigate potential contamination, considering all of these objects exhibit distorted morphologies, many of which were initially classified as either FRIs or RDs by the majority classification vote highlighted in Fig. \ref{fig:matrix}. Fig. \ref{fig:agn_selection} displays the distribution of our sample overlaid on the D24 sample. This plot illustrates the cuts we implemented in flux, luminosity, and size during our visual morphological classification process. Our RLAGN subset (in red) occupies the upper right part of the plot, indicating higher radio luminosities and larger flux densities compared to the broader sample from D24. This suggests that our selected RLAGN are more powerful and extended sources, though this trend is influenced by selection bias due to the flux and size limits applied in this study. The inset showing the redshift distribution reveals that the majority of our RLAGN are at moderate redshifts, peaking around $z \sim 0.25$, which is advantageous for detailed follow-up studies.

\subsection{Emission-line classifications of HERGs and LERGs}\label{sec:herg_lerg}
\begin{figure}
    \centering
    \includegraphics[width=\columnwidth]{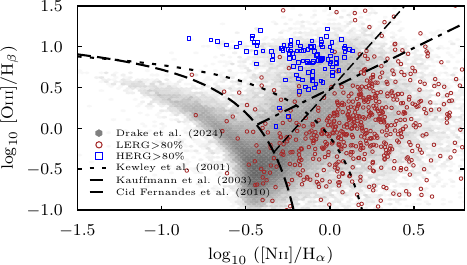}
    \caption{The BPT diagram highlighting the positions of objects in our sample, spectroscopically classified as HERG (squares) and LERG (circles) using the 80 per cent threshold, overlaid on the D24 sample (hexagons)}.
    \label{fig:bpt}
\end{figure}

Table \ref{tab:lherg} presents the statistics for our sample, categorized into HERG and LERG classes according to the methods outlined by D24 (summarized in Sec. \ref{sec:data}). All objects in our sample are classified as radio-excess AGN with a confidence level greater than 90 per cent. Within this broader classification, we identified 108 objects as HERGs and 1256 as LERGs, based on their optical emission lines, with a confidence level exceeding 80 per cent. The distribution of these objects on the BPT diagram is shown in Fig. \ref{fig:bpt}.

\begin{table}
    \centering
    \caption{A summary of our objects, spectroscopically separated into HERG and LERG categories using the BPT diagram with an 80 per cent threshold.}
    \begin{tabular}{c|c|c|c}
     \hline\hline
     Morphology &LERG & HERG & \# of objects \\
     \hline
     FRI  & 371 & 2 & 373 \\
     FRII & 404 & 104 & 508 \\
     WAT  & 203 & 1 & 204\\
     NAT  & 66 & 0 & 66 \\
     HT  & 129 & 0 & 129\\
     RD  & 83 & 1 & 84\\
     \hline
     Total  & 1256 & 108 & 1364 \\
     \hline\hline
    \end{tabular}
    \label{tab:lherg}
\end{table}

\begin{figure}
    \centering
    \includegraphics[width=\columnwidth]{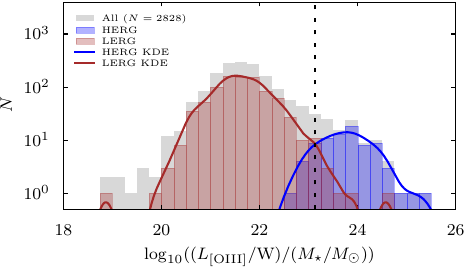}
    \caption{An adaptation of the D24 HERG/LERG classification criteria utilizing the specific [O{\scriptsize{III}}] emission-line luminosity for radio-excess AGN with available [O{\scriptsize{III}}] and stellar mass measurements. Objects with log$_{10}~(L\rm{_{[OIII]}}$$/M_{\star})>23.14$ are classified as HERGs. Those below this threshold are classified as LERGs.}
    \label{fig:o3m_lherg}
\end{figure}

To identify additional HERGs and LERGs that were missed by the 80 per cent threshold, Fig. \ref{fig:o3m_lherg} presents the results of applying the specific [O{\scriptsize{III}}] emission-line luminosity method, as outlined in D24. After filtering out the HERGs and LERGs selected with a confidence level greater than 80 per cent, we applied a threshold of log$_{10}~(L\rm{_{[OIII]}}$$/M_{\star})=23.14$ to the remaining sample. This resulted in the classification of 26 additional HERGs and 719 additional LERGs for objects with available [O{\scriptsize{III}}] and stellar mass measurements.

\begin{table}
    \centering
    \caption{The selection of additional HERGs and LERGs using the specific [O{\scriptsize{III}}] emission-line luminosity method for objects with less than 80 per cent confidence in emission-line classifications.}
    \begin{tabular}{c|c|c|c}
     \hline\hline
     Morphology & LERG & HERG & \# of objects \\
     \hline
     FRI  & 225 & 4 & 229 \\
     FRII & 257 & 17 & 274 \\
     WAT  & 108 & 1 & 109\\
     NAT  & 38 & 0 & 38 \\
     HT  & 59 & 1 & 60\\
     RD  & 32 & 3 & 35\\
     \hline
     Total  & 719 & 26 & 745 \\
     \hline\hline
    \end{tabular}
    \label{tab:sum_morph3}
\end{table}

The remaining 719 sources in our sample are radio-excess objects that could not be classified using the BPT diagram or their specific [O{\scriptsize{III}}] emission-line luminosity. However, their radio morphologies, which show clear features typical of RLAGN, strongly suggest that the radio emission originates from AGN activity. Considering that most LERGs exhibit weak or absent ionization emission lines like [O{\scriptsize{III}}] in their optical spectra, making them difficult to classify with these methods, we classified these sources as LERGs. This resulted in a sample of 134 HERGs and 2694 LERGs. The statistics for each morphological and emission-line class are provided in Table \ref{tab:final_class}. Overall, our final sample is dominated by LERGs, with HERGs representing a minority, as expected for RLAGN at lower redshifts \citep{1979MNRAS.188..111H,2020NewAR..8801539H}. We discuss the distribution of morphological and optical emission-line properties in the next section.

\begin{table}
    \centering
    \caption{This table presents the final morphological sample, classified into HERGs and LERGs using the methods described above.}
    \begin{tabular}{c|c|c|c}
    \hline \hline
     Morphology & HERG & LERG & \# of objects  \\
     \hline
     FRI  & 6 & 806 & 812\\
     FRII & 121 & 839 & 960\\
     WAT & 2 & 471 & 473\\
     NAT & 0 & 158 & 158\\
     HT &  1 & 271 & 272\\
     RD & 4 & 149 & 153\\
     \hline
     Total & 134 & 2694  & 2828\\
     \hline\hline
    \end{tabular}
    \label{tab:final_class}
\end{table}

\section{Results} \label{sec:results}

\subsection{The power/linear size}\label{sec:pd}
The power/linear size (P-D) diagram \citep{1982IAUS...97...21B,2018MNRAS.475.2768H} is a powerful diagnostic tool for classifying RLAGN and tracing their evolutionary stages.

\begin{figure*}
    \centering
    \includegraphics[width=2\columnwidth]{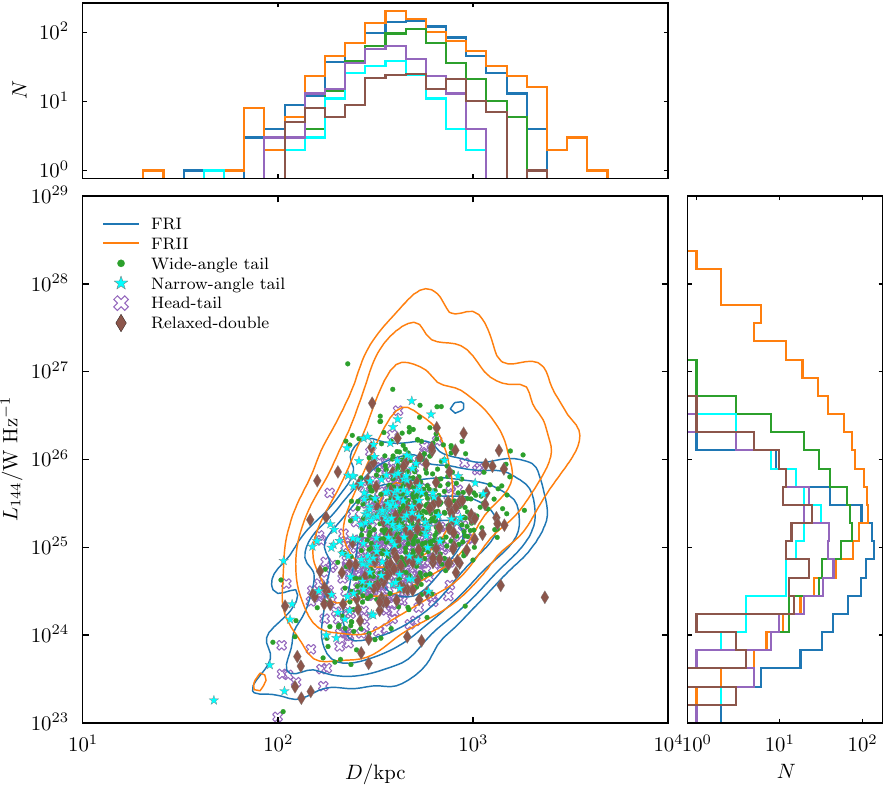}
    \caption{P-D diagram showing the distribution of visually classified morphological sources in our sample. Histograms for size and luminosity are included to highlight the range covered by each object.}
    \label{fig:pd_diagram}
\end{figure*}

\begin{figure}
    \centering
    \includegraphics[width=\columnwidth]{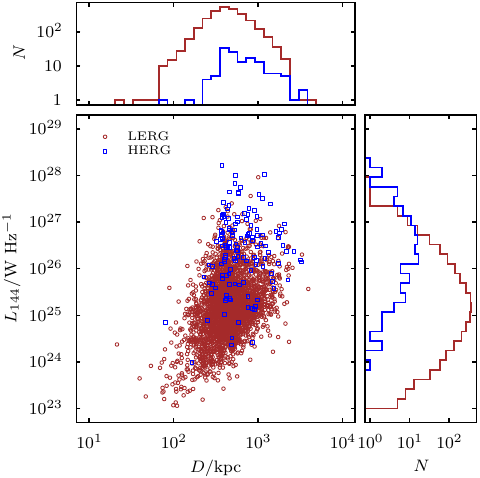}
    \caption{The distribution of LERGs (shown as brown open circles) and HERGs (blue open squares) in P-D space. This histograms highlight the coverage in size and luminosity for LERG and HERG in our sample.}
    \label{fig:pd_diagram2}
\end{figure}

Fig. \ref{fig:pd_diagram} highlights the diversity among RLAGN in our sample in terms of radio morphology. This plot include 2828 objects (see Table \ref{tab:final_class} for the final sample breakdown). We observe that FRI sources (blue contour lines) generally exhibit lower luminosities (804 out of 812), typically below $\sim$1 $\times~10^{26}~\rm{W~Hz^{-1}}$ at 144 MHz -- a threshold we use to differentiate more luminous sources from those below it, following the traditional FRI/FRII luminosity break \citep{1974MNRAS.167P..31F,1996AJ....112....9L} -- and have smaller sizes compared to FRII sources (orange contour lines). Thus, most FRIs occupy the lower half of the P-D space, with only a small fraction (8 out of 812) exceeding the FRI/FRII luminosity break. The largest FRI in our sample spans $\sim$2205 kpc, while the largest FRII spans $\sim$3948 kpc. FRI sizes are measured from the largest angular extent of the radio emission, while FRII sizes are based on the full extent of the radio lobes. However, it should be noted that size measurements for FRIs are ill-defined due to surface brightness limitations in the current survey. In contrast, the high-luminosity end of the P-D diagram is dominated by FRIIs (294 out of 380 total sources in this region), suggesting differences in energy output, environment, or central engine characteristics between these two classes. 

We also identify a significant number of low-luminosity FRIIs (666 out of 960, approximately 69 per cent), which is consistent with previous observations of low-luminosity FRIIs \citep[e.g.][]{1994ASPC...54..201L,2017MNRAS.466.4346M,2019MNRAS.488.2701M}. However, we observe an increase in low-luminosity FRIIs compared to the 51 per cent reported by \citet{2019MNRAS.488.2701M} and Clews et al, \textit{in prep}. These low-luminosity FRIIs might represent a transitional phase or a distinct evolutionary path compared to their more powerful counterparts, potentially located in environments with lower gas densities \citep{2019MNRAS.488.2701M}, which helps the jet to stay focused and supersonic. Additionally, a decrease in central engine power over cosmic time may alter the balance between jet dynamics and environmental resistance, contributing to their observed characteristics.

WATs (green points), NATs (cyan stars), HTs (purple open crosses), and RDs (brown diamonds) exhibit a wide range of sizes and luminosities, overlapping with both FRIs and FRIIs. The majority of WATs, NATs, and HTs (423/473, 146/158, and 266/272, respectively) are found in the FRI/II region of the lower P-D space, with a smaller fraction (50/473 WATs, 12/158 NATs, and 6/272 HTs) exceeding $\sim$$1~\times~10^{26}~\rm{W~Hz^{-1}}$ at 144 MHz. The overlap of, for example, WATs in the low-luminosity range with FRIIs supports the idea that these objects are essentially `failed' FRII sources \citep[e.g.][]{2004MNRAS.349..560H}, whose morphology and radio power are influenced by varying environmental conditions.

Specifically, WATs, often found in central galaxy clusters \citep[e.g.][and references therein]{2023Galax..11...67O}, display a diverse distribution in the P-D diagram, reflecting complex jet-ICM interactions that can deflect their jets, resulting in distinctive morphologies and variability in size and luminosity. Similarly, NATs and HTs, typically associated with galaxies moving through dense cluster environments, cover a wide range in P-D space, indicating that ram pressure stripping and ICM interactions significantly shape their structure and radio power. The spread in their observed properties is likely due to the varying environments and velocities within clusters.

RDs seem to occupy a transitional stage in the P-D diagram. The majority (143 out of 153) fall below $\sim$$1~\times~10^{26}~\rm{W~Hz^{-1}}$ at 144 MHz, with a minority (10 out of 153) above this threshold. Their position, coinciding with low-luminosity FRIIs, suggests they might be evolving from FRII-like morphologies as they lose energy and become less luminous. This transition could be driven by a combination of declining jet power, central engine changes, and environmental factors, such as expansion in lower-density regions, leading to their relaxed appearance \citep{Leah93}.

In total, we have 2448 low-luminosity sources and 380 high-luminosity sources in these plots. The substantial overlap of these sources with both FRI and FRII classes further emphasizes that radio morphology and luminosity are not solely determined by the central engine but are also significantly influenced by external factors such as the density of the surrounding medium, dynamics within galaxy clusters, and the evolutionary stage of the AGN. This complexity underscores the importance of considering environmental effects in the morphological classification of RLAGN.

In Fig. \ref{fig:pd_diagram2}, we highlight the distribution of our sample, separating  high-excitation and low-excitation radio galaxies (HERG and LERG, respectively). As expected, HERGs (squares) are generally more luminous (96 out of 134) and span a wide range of sizes. Conversely, LERGs (circles) typically exhibit lower luminosities (2410 out of 2694) and cover a broader size range, suggesting different accretion modes. This distinction in luminosity and size distribution reinforces the idea that HERGs are powered by RE accretion onto rapidly spinning black holes, while LERGs are associated with RI accretion, possibly from hot gas reservoirs \citep[e.g.][]{1989ApJ...336..681B,2007MNRAS.376.1849H}. The broader size range observed for LERGs may indicate different evolutionary paths, potentially influenced by environmental factors or longer lifetimes in less dense surroundings \citep{2018MNRAS.475.2768H}. The histograms along the axes further emphasize these differences, with the luminosity distribution of HERGs peaking at higher values compared to LERGs, and the size distribution of LERGs showing a broader spread.

We also note that the majority of HERGs are classified as FRIIs (121 out of 134), with 92 of these 121 FRII HERGs above the FRI/FRII luminosity break and 29 below it. This observation aligns with the suggestion that while most HERGs are FRIIs, not all FRIIs are HERGs \citep[e.g.][]{2006MNRAS.370.1893H,2010A&A...509A...6B}; we expand on this point below, with further discussion provided in Sec. \ref{sec:fr2_sources}. A small subset of HERGs is classified as FRIs (6 out of 134), with two of these FRI HERGs above the luminosity break and four below, consistent with previous findings that FRI HERGs are rare \citep[e.g.][]{2022MNRAS.511.3250M}. Additionally, our HERG sample includes two WATs (one above and one below the luminosity break), one HT (below the break), and four RDs $-$ three of which are below the break, with one RD above it.

The LERG sample comprises 284 out of 2694 objects that are above the traditional FRI/FRII luminosity break. These include 202 FRIIs, 6 FRIs, 49 WATs, 12 NATs, 6 HTs, and 9 RDs. The remaining 2410 objects, categorized as 800 FRIs, 637 FRIIs, 422 WATs, 146 NATs, 265 HTs, and 140 RDs, fall below this luminosity threshold. Although our final sample is dominated by FRII sources (Table \ref{tab:final_class}), which is expected given the size cut used in the classification, we observe that most FRIs in our sample are LERGs, with the majority occupying the lower half of the P-D space, as expected. Furthermore, as noted earlier, a significant number of FRII LERGs populate the low-luminosity space, with a substantial portion also found above the luminosity break. This distribution emphasizes that FRII sources can be either HERGs or LERGs depending on their evolutionary path. Notably, it seems plausible that some FRII LERGs appear to evolve into RDs as their jets cease activity, leading to a transition into less active states. The presence of a notable population of RD LERGs with FRII-like structures suggests that, once the jet turns off, these sources expand and relax into lower-energy configurations characteristic of remnant radio galaxies.

Although most of our objects occupy the expected regions on the P-D diagram, it is important to note that the size cuts used in our classification have led to the exclusion of a significant population of sources. Many of these are compact objects that can only be resolved at higher angular resolution \citep[e.g.,][]{2024MNRAS.529.1472C}, and they would occupy the left side of the P-D diagram. Similarly, on the right side, we are missing a population of physically-large low-luminosity RLAGN, likely due to limitations in surface brightness sensitivity in current surveys \citep{2019A&A...622A..12H}. Despite these gaps, our analysis using the P-D diagram supports the robustness of our sample classifications in both morphology and optical emission-line class, which is crucial for the subsequent sections.

\subsection{WISE colours}\label{sec:wise}
\begin{figure}
    \centering
    \includegraphics[width=\columnwidth]{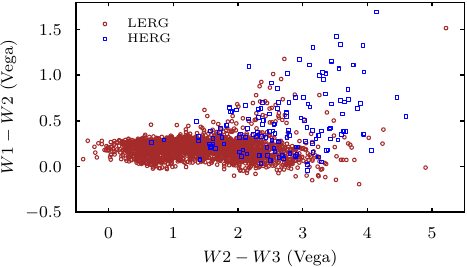}
    \caption{The position of HERGs and LERGs in WISE colour-colour plots. For a discussion of upper limits, refer to the caption in Fig. \ref{fig:wise_cc2}.}
    \label{fig:wise_cc}
\end{figure}

\begin{figure*}
    \centering
    \includegraphics[width=2\columnwidth]{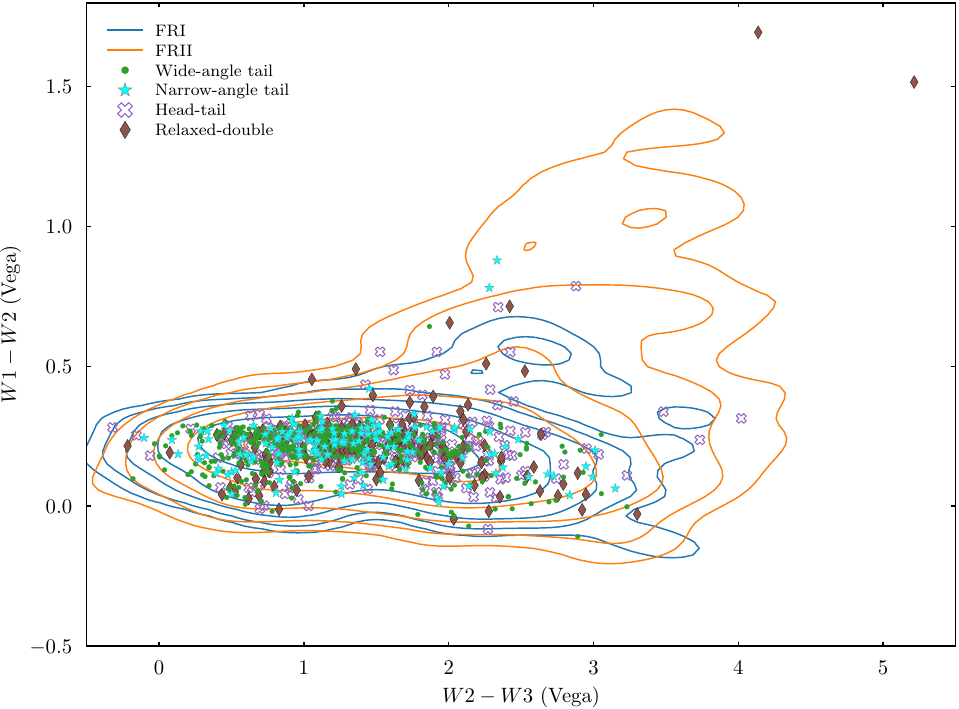}
    \caption{The distribution of various morphological classes within our sample in the WISE colour-colour space. Approximately 50 per cent of our sources have upper limits in W3. To avoid clutter, we have not included left arrows to indicate that their true positions might be farther to the left than depicted.}
    \label{fig:wise_cc2}
\end{figure*}

\begin{figure*}
    \includegraphics[width=\columnwidth]{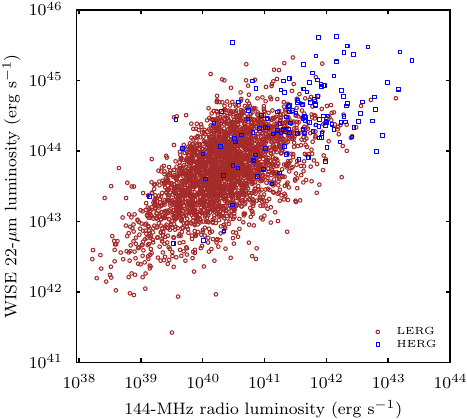}\hspace{5mm}
    \includegraphics[width=\columnwidth]{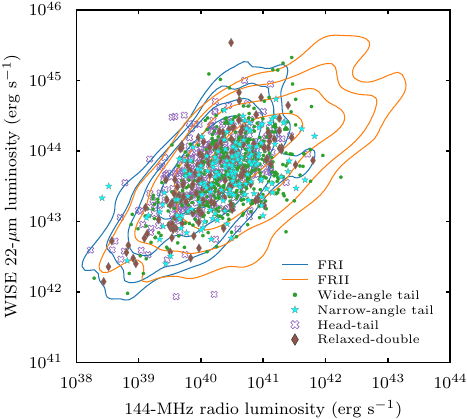}
    \caption{The figure shows the distribution of optical emission-line classifications (left panel) and morphological classifications (right panel) on the WISE $W4$ luminosity versus radio luminosity plane. Of the 2828 sources in our sample, only 766 have $W4$ detections, while the rest are represented as upper limits.}
    \label{fig:wise_lum}
\end{figure*}

The {\it Wide-field Infrared Survey Explorer} \citep[WISE;][]{2010AJ....140.1868W} plots shown in Fig. \ref{fig:wise_cc} $-$ \ref{fig:wise_lum} are a powerful tool for understanding RLAGN based on their mid-infrared (IR) properties \citep[e.g.][Hardcastle et al., {\it submitted}]{2014MNRAS.438.1149G,2016MNRAS.462.2631M}. The WISE $W1$ band at 3.4 $\mu$m captures a mix of stellar emission from the host galaxy and AGN-related components such as hot dust and synchrotron emission. This band can be used as a proxy for estimating host galaxy masses and consequently black hole masses. The $W2$ band at 4.6 $\mu$m reflects similar contributions but with a stronger influence from the torus of the AGN. The $W3$ band at 12 $\mu$m, which is highly sensitive both to warm dust from star forming galaxies and to the torus, plays a crucial role in the separation seen in these plots. While the $W4$ band at 22 $\mu$m also captures a mix of radiation from SFGs and AGNs, it is more heavily influenced by the AGN radiation, particularly from the cooler dust in the outer regions of the AGN torus. The distribution of RLAGN in these colour spaces therefore offers valuable insights into the nature of their host galaxies, which we explore further below.

Fig. \ref{fig:wise_cc} presents the WISE colour-colour diagram highlighting the distribution of our sample in optical emission-line classifications. As expected, a majority of our HERGs occupy regions with higher values of $W2-W3>2$, while LERGs are clustered at lower values $W1-W2<0.5$ and $W2-W3<2$, reflecting stronger and weaker mid-IR emission, respectively. This means that HERGs typically exhibit more substantial amounts of warm dust than LERGs \citep[e.g.][]{2005MNRAS.362....9B,2018MNRAS.475.3429W}, as their higher levels of SF and AGN activity generate more intense radiation, which heats the surrounding dust \citep[e.g.][]{2010ApJ...722.1333D,2012ApJ...745..172D}. However, we see some overlap between the two classifications, as expected, given that nearly half of the sources in our sample lack detections in the $W3$ band and are represented by upper limits. Many of the upper limits in $W3$ are from LERGs (1396 out of 2694), with a minority coming from HERGs (27 out of 134).

If we adapt Fig. \ref{fig:wise_cc} to the morphological classifications in our sample, we find that a significant fraction of FRI, FRII, WAT, NAT, HT, and RD sources occupy the expected regions (Fig. \ref{fig:wise_cc2}), predominantly associated with massive elliptical hosts, with many classified as LERGs. Additionally, the objects located in the HERG region of Fig. \ref{fig:wise_cc} are mostly high-luminosity FRII sources, as noted in the previous section. The differences between FRII HERG and FRII LERG environments suggest distinct host galaxy properties despite their similar morphologies. FRII HERG hosts may exhibit substantial amounts of warm dust, as earlier noted for the general HERG population, while FRII LERG hosts are likely associated with more quiescent galaxies, consistent with the general characteristics of the LERG population.

Recent studies \citep[e.g.,][Hardcastle et al., {\it submitted}]{2014MNRAS.438.1149G} have demonstrated a correlation between low-frequency radio luminosity and mid-IR luminosity, showing a clear separation between HERGs and LERGs around $10^{43}$ erg s$^{-1}$ in the WISE $W4$ band. This correlation provides a basis for inferring the relationship between optical emission-line classification and morphological classification within our sample. To investigate this, we calculated $W4$ luminosities for our sources by first estimating the mean spectral index between $W3$ and $W4$. WISE Vega magnitudes were then converted to flux densities, applying a $K$-correction term to derive $W4$ luminosities. Additionally, radio luminosities in W Hz$^{-1}$ were converted to units of erg s$^{-1}$ to ensure consistency with the mid-IR luminosities, enabling a direct comparison between the two.

Fig. \ref{fig:wise_lum} is a plot of $W4$ luminosity versus radio luminosity. The majority of HERGs (left panel) exhibit higher $W4$ luminosities, consistent with the findings of \citet{2014MNRAS.438.1149G}. However, a significant fraction of LERGs also exceed this $W4$ luminosity threshold, though the majority of their values are upper limits. Notably, only about 27 per cent of the sources in this plot have detections in $W4$, which naturally leads to considerable overlap between the classifications.

On the right panel of Fig. \ref{fig:wise_lum}, FRIIs display a wide range of luminosities, spanning from low to high values. This distribution aligns with the presence of numerous low-luminosity sources at 144 MHz frequencies \citep{2019MNRAS.488.2701M}. Specifically, FRII HERGs occupy regions above $10^{44}$ erg s$^{-1}$ in $W4$ luminosity and $10^{41}$ erg s$^{-1}$ in radio luminosity. In contrast, FRI and WAT sources exhibit scattering on this plane, with many FRIs seen below the $W4-L_{\mathrm{144}}$ threshold. Meanwhile, other morphological classes tend to cluster along the `main sequence', suggesting a more uniform distribution.

These results indicate that distinctions in radio morphology, traditionally established through radio observations, are also evident in mid-IR observations. However, caution is necessary when classifying AGN using these methods, as overlaps between classes are significant. As shown in the right panel of Fig. \ref{fig:wise_lum}, only FRII HERGs can be classified with reasonable precision using this approach. 

The results of this study complement previous research \citep[e.g.][]{2012ApJ...753...30S,2012MNRAS.426.3271M,2014MNRAS.438.1149G,2016MNRAS.462.2631M,2018MNRAS.480..707P,2019MNRAS.488.2701M,2019A&A...622A..12H}, highlighting the value of mid-IR observations in enhancing radio and optical studies \citep{2013ApJ...772...26A}. This multi-wavelength approach provides deeper insights into the nature of RLAGN and their surrounding environments.

\subsection{Host galaxy mass}\label{sec:host_mass}
In the previous section, we discussed the environments of our sources. Here, we shift focus to examine their host galaxy masses using mass estimates from \citet{2023A&A...678A.151H}. For this analysis, we only consider objects with reliable mass estimates above the threshold of $\log_{10}(M_{\star}/M_\odot)>10^{8.5}$, as values below this are considered unreliable due to uncertainties in photometric redshift estimates. This selection leaves us with 2185 out of 2828 objects, which we analyze below.

\begin{figure}
    \centering
    \includegraphics[width=\columnwidth]{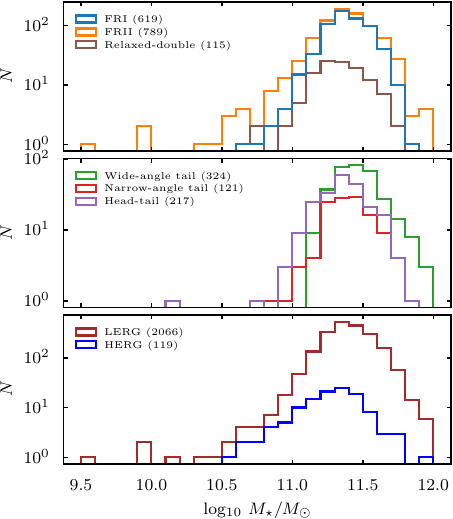}
    \caption{Distribution of host galaxy mass categorized by morphology (top and middle) and emission-line class (bottom). The legend provides a breakdown of each classification with reliable host mass estimates.}
    \label{fig:steler_mas}
\end{figure}

Fig. \ref{fig:steler_mas} provides insights into the distribution of host galaxy masses for various RLAGN populations in our sample, classified by their radio morphology (top and middle plots) and excitation type (bottom plot). As shown in the top plot, FRI sources tend to have slightly higher host masses on average compared to FRII sources, as expected, though the difference is not pronounced at higher stellar masses. However, we notice a tail of low-mass FRIIs below log$_{10}(M_{\star}/M_{\odot})<10.5$. This result has also been reported previously \citep[e.g.][]{2019MNRAS.488.2701M,2022MNRAS.511.3250M}, and likely reflects a selection bias, given the significant population of low-luminosity FRIIs in our sample as highlighted in Sec. \ref{sec:pd}. FRIIs in our sample therefore span a wide range of host galaxy masses, from moderately massive to very massive systems, showing that these objects are found across a broad host mass spectrum. We also observe that RDs follow the same trend as FRIIs, which aligns with previous suggestions that RDs are fading FRIIs whose jet activity has been `turned off'.

The WATs in our sample on average show host galaxies that are on the more massive end compared to NATs and HTs. This is consistent with previous findings \citep[e.g.][]{2004MNRAS.349..560H} where these objects are expected to be in the central regions of galaxy clusters and so would naturally be associated with massive galaxies. A large fraction of NATs however are biased towards the lower-mass end of the plot implying that these sources reside in lower-mass galaxies. This is also expected as NATs are often associated with galaxies lying on the outskirts of groups or clusters.

As noted in Sec. \ref{sec:into}, HERGs are generally observed to reside in less massive galaxies than LERGs \citep[e.g.,][]{2005MNRAS.362....9B,2009ApJ...696...24S,2012MNRAS.421.1569B,2013MNRAS.430.3086G,2022MNRAS.513.3742K}, often in environments with abundant cold gas that can fuel star formation. In contrast, LERGs are predominantly found in more massive elliptical galaxies, characterised by older stellar populations and less gas available for accretion compared HERGs. The bottom panel of Fig. \ref{fig:steler_mas} confirms this trend, showing that LERG host galaxies are generally more massive than those of HERGs. We also see that LERGs display a broader mass distribution, including a tail of lower-mass sources. This low-mass subset belongs to the low-luminosity FRII category discussed earlier and occupies the SFG region in the WISE colour-colour diagram \citep[Sec. \ref{sec:wise}]{2016MNRAS.462.2631M}. Additional optical data, such as from the WEAVE-LOFAR spectroscopic survey \citep{2016sf2a.conf..271S}, is needed to confirm the fueling mechanism for this low-mass category.

\subsection{Core prominence}\label{sec:core_pro}
The core prominence ($P_c$), defined as the ratio of core to total radio emission, is a key parameter in understanding the physical processes occurring in these RLAGN. It offers insights into the orientation of radio jets relative to our line of sight \citep[e.g.][]{1982MNRAS.200.1067O}, the activity state of the AGN, and the underlying mechanisms driving jet production and evolution \citep[e.g.][]{1995MNRAS.276.1215S,2015A&A...576A..38B}. A key advantage of the 144-MHz radio observations used in this study is their ability to detect older radio populations \citep[e.g.][]{2016MNRAS.462.1910H}, which are often challenging to observe at higher frequencies. This allows us to use the core prominence to explore the different morphological and emission-line classifications within our sample, as well as to identify and classify other AGN populations, such as remnant and restarted candidate sources \citep[e.g.][]{2017A&A...606A..98B,2019A&A...622A..13M,2024A&A...691A.287N}.

\begin{figure}
    \centering
    \includegraphics[width=\columnwidth]{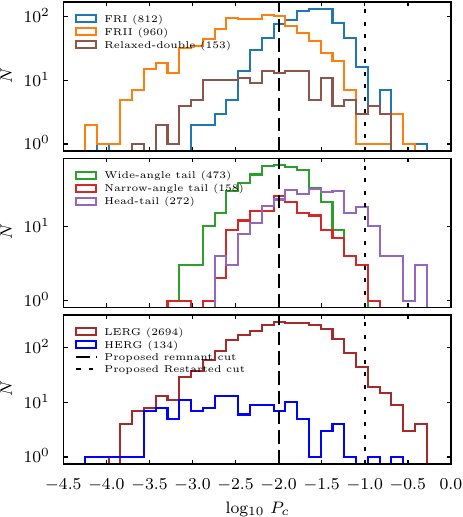}
    \caption{Core prominence ($P_c$) distribution for the various morphological classes (top and middle) and emission-line class (bottom) in our sample. Upper limits are also included on these plots. We additionally included proposed cuts used in the selection of remnant and restarted candidates \citep[e.g.][]{2017A&A...606A..98B,2021A&A...653A.110J}.}
    \label{fig:core_prom}
\end{figure}

To calculate the $P_c$ parameter, we first obtained VLASS images \citep{2020PASP..132c5001L,2021ApJS..255...30G} for all our objects and used the optical RA and DEC coordinates to identify the brightest pixel within each VLASS field. Operating at a frequency of 3 GHz (angular resolution of 2.5 arcsec), VLASS provides higher-frequency radio data compared to the 144-MHz data from LOFAR. For each source, we recorded the local root mean square (RMS) noise. If the measured signal-to-noise ratio fell below the 3$\sigma$ threshold, as was the case for 216 objects in our sample, we replaced the flux value with an upper limit equal to the 3$\sigma$ value to account for non-detections. We then compared these VLASS core flux measurements to the LOFAR total flux at 144 MHz to calculate $P_c$ for each source. This approach allowed us to assess the AGN core contribution at different frequencies.

In general, core prominence values for RLAGN fall within the range $-4<$~log$_{10}~P_c<0$ \citep[e.g.][]{1998MNRAS.296..445H,2008MNRAS.390..595M,2014MNRAS.437.3405L}, with FRIs generally showing higher $P_c$ values than FRIIs. This trend is reflected in our sample (top panel of Fig. \ref{fig:core_prom}), where on average, FRIs seem to have higher $P_c$ values compared to FRIIs. This result means that a greater contribution in FRIs comes from the core emission relative to the total radio output. This behavior may be attributed to a potential restarting phase, differences in their jet formation processes, orientations, and other underlying physical conditions. For instance, FRIs having lower radio luminosities than FRIIs on average can result in an increased content of non-radiating particles for the same jet power \citep[e.g.][]{2018MNRAS.476.1614C}, which can contribute to their high core prominences. Additionally, FRI jets being slower and less collimated \citep[e.g.][]{2014MNRAS.437.3405L} suggests that dissipation may occur on smaller scales, in which case can lead to a higher fraction of the emission being concentrated in the core region. FRIIs, however, are characterised by their more powerful and well-collimated jets, exhibiting significant extended lobes, which may dilute the core contribution in the total flux. 

We also observe that WATs and NATs in our sample have intermediate core prominences between FRIs and FRIIs, reflecting the mixed nature of their structures. The significant bulk deceleration of NAT jets on kpc scales, due to interactions with the ICM, as also observed in FRIs \citep[e.g.][]{2014MNRAS.437.3405L}, likely explains the behaviour in these sources as well. This interaction can slow the jets and redistribute energy between the core and extended lobes, resulting in intermediate core prominences seen in Fig. \ref{fig:core_prom}. HTs in our sample, however, exhibit higher core prominences on average compared to WATs and NATs, with values overlapping those of FRIs on the more core-dominated side of the distribution. However, we do not have sufficient resolution to distinguish jet bases in FRIs, WATs, NATs, and HTs from the flat spectrum core, which might contribute to the high core prominence observed in these sources. Additionally, higher $P_c$ values in HTs can be attributed to the more concentrated energy output in the core region, likely due to less disruption of the central jets and stronger collimation. Higher core prominence values in some HTs may also be attributed to relativistic beaming, where the one-sided radio morphology, characteristic of HTs, could indicate strongly beamed sources \citep[e.g.][]{2016ApJ...830...82M}.

It is unsurprising to observe a population of RDs in our sample exhibiting intermediate $P_c$ values, overlapping with both FRIs and FRIIs. A fraction of these sources have detectable (albeit faint) radio cores, which explains the overlap of their core prominence values with those of FRIIs. This can be attributed to the aging and deceleration of the jets, particularly in the remnant phase, where the lobes fade more rapidly than the core, leading to a higher relative core contribution compared to the extended emission, thereby elevating $P_c$ values. Additionally, selection bias may play a role, as structures with FRII-like morphologies but lacking prominent hotspots were classified as RDs instead of FRIIs. The higher core values could also indicate renewed AGN activity (restarted jets; see Sec. \ref{sec:other_classes}), where fresh energy injection brightens the core even as the outer lobes fade, further increasing the core prominence.

In the bottom panel of Fig. \ref{fig:core_prom} we present the distribution of our sample separated into HERGs and LERGs. We observe a continuous distribution of $P_c$ values within the LERG sample, with core prominences peaking around log$_{10}~P_c\sim-2$. Around 54 per cent of LERGs are above this threshold, with FRIs dominating this area followed by HTs in comparison to their total populations in the entire sample. Additionally, we note a significant presence of more core-dominated FRIIs in this region, as also highlighted in the morphological distribution. A minority of these FRIIs show signatures of restarted AGN activity (see Sec. \ref{sec:other_classes}), which is consistent with the characteristics of more core-dominated FRIIs. Conversely, around 46 per cent of LERGs fall below this threshold, which coincidentally aligns with the proposed cut for selecting remnant source candidates \citep[e.g.][]{2021A&A...653A.110J}. We will revisit this point shortly, noting that FRII LERGs dominate this region. 

\begin{figure}
    \centering
    \includegraphics[width=\columnwidth]{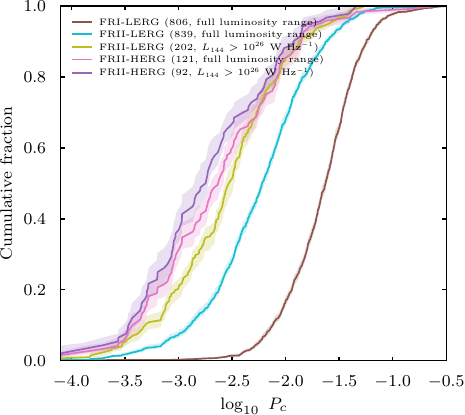}
    \caption{The distribution of core prominence within the FRII sample, separated into LERG and HERG. The analysis includes the full luminosity range and a subset of high luminosity FRIIs, as described in the text. The FRI LERG sample is included for comparison. The number of objects for each category is included in the legend. The shaded regions represent the 1$\sigma$ errors, estimated using the bootstrapping procedure.}
    \label{fig:fr2_only_R}
\end{figure}

We expect HERGs to show higher average values of $P_c$ compared to LERGs due to their active central regions. However, this expectation is not met in our sample (Fig. \ref{fig:core_prom}). Approximately 83 per cent of HERGs fall below log$_{10}~P_c\sim-2$, and those above this threshold display signs of restarted activity, with FRII sources dominating this category (see Table \ref{tab:other_classes}). To further investigate the low core prominence observed in HERGs compared to LERGs, we analyse the FRII sample which includes a large fraction of both HERGs and LERGs. This also provides an opportunity to test the hypothesis that FRII LERGs are `switched off' HERGs \citep[e.g.][]{2016A&ARv..24...10T,2020MNRAS.494.2053P,2020MNRAS.493.4355M,2022MNRAS.511.3250M}. If FRII LERGs represent a transitioned population of FRII HERGs, we would expect their core prominence distribution to be shifted to lower values than that of FRII HERGs, due to decreased core activity. However, as shown in Fig. \ref{fig:fr2_only_R}, FRII LERGs exhibit higher core prominence values than FRII HERGs, suggesting that these two populations are distinct in this context. A similar trend is observed when considering only high luminosity FRII LERGs and HERGs (i.e. above the nominal threshold discussed in this work). This conclusion is reinforced by the Mann-Whitney U test, which reveals significant differences in their distributions, with $p$-value of $1.5 \times 10^{-11}$ and 0.01, respectively. These results suggest a fundamental distinction between FRII-LERGs and their HERG counterparts, positioning FRII-LERGs as an intermediate population between FRI-LERGs and FRII-HERGs. A further discussion on the FRII sample is provided in Sec. \ref{sec:fr2_sources}.  

\begin{figure}
    \centering
    \includegraphics[width=\columnwidth]{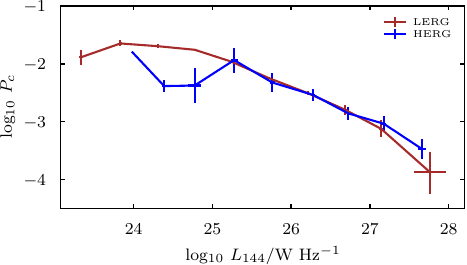}
    \caption{The distribution of core prominence as a function of radio luminosity. The points, binned by luminosity with a bin size of $\Delta$log$_{10}~L_{144}/\rm{W~Hz^{-1}} = 0.5$, represent the mean values with bootstrapped errors, estimated from the 68th percentile confidence interval.}
    \label{fig:lum_vs_R}
\end{figure}

For the broader LERG and HERG populations, previous studies have demonstrated a strong dependence of core prominence on radio luminosity for both groups. Specifically, HERGs, which are typically higher radio-powered sources, tend to exhibit lower core prominence values compared to LERGs \citep[e.g.][]{2008MNRAS.390..595M}. This trend suggests that, as radio power increases, the core prominence of HERGs diminishes relative to LERGs. This can be attributed to the fact that more powerful radio galaxies often have more extended radio structures, causing a greater proportion of emission to be spread across larger regions of the source. As a result, the contribution from the central active region is lower relative to the total emission, leading to a decrease in core prominence. However, in our sample, we find no significant difference between HERGs and LERGs. Instead, both populations show a decrease in core prominence with increasing radio luminosity, as highlighted in Fig. \ref{fig:lum_vs_R}.

In addition to the discussion above, we investigated the application of proposed selection criteria for identifying remnant and restarted RLAGN candidates, using cuts at log$_{10}~P_c<-2$ and log$_{10}~P_c>-1$, respectively \citep[e.g.][]{2017A&A...606A..98B,2021A&A...653A.110J}. However, adopting these criteria would classify nearly half of our objects as remnant or restarting sources, which seems implausible. The histograms in Fig. \ref{fig:core_prom} reveal a more continuous distribution of core prominences, rather than distinct populations, suggesting that core prominence may not provide a clear-cut distinction between remnant and active sources. This continuous distribution could reflect the complex and gradual evolutionary pathways of RLAGN, where transitions between different phases (active, remnant, or restarted) are not sharply defined, but instead, overlap significantly. Therefore, we adopt a visual classification in this work, as was used by e.g. \citet{2012ApJS..199...27S}, \citet{2018MNRAS.475.4557M}, and \citet{2019A&A...622A..13M} in selecting these objects. 

In principle, remnant sources should show no signs of central AGN activity and can be identified by the absence of a radio core at all frequencies, with only extended emission at lower frequencies, often taking the form of amorphous structures \citep[e.g.][]{2018MNRAS.475.4557M}. Following the remnant phase is the restarted phase, sometimes characterised by a steep-spectrum radio core \citep[e.g.][]{1996MNRAS.278....1A}, and radio emission that mirrors past AGN activity. In objects with FRII-like structures, this may manifest as small double lobes aligned with older, larger lobes, forming the so-called double-double radio galaxies \citep[DDRGs; e.g.][]{2000MNRAS.315..371S, 2019A&A...622A..13M,2024arXiv240813607D}. Therefore, restarted candidates can generally be selected by noting large-scale amorphous structures at lower frequencies and bright radio cores across all (or at higher) frequencies. We will discuss this further in Sec. \ref{sec:remt_rest}.

\section{Discussion}\label{sec:discusion}
\begin{figure}
    \centering
    \includegraphics[width=\columnwidth]{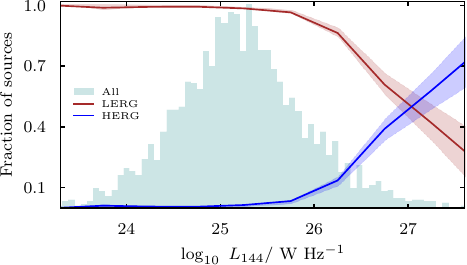}
    \caption{Fraction of low-excitation and high-excitation radio galaxies in our sample as a function of radio luminosity. Shaded regions represent Binomial-based errors, indicating the uncertainty in the fraction of sources within each bin of size $\Delta$log$_{10}~L_{144}/\rm{W~Hz^{-1}} = 0.5$.}
    \label{fig:frac_lerg_herg}
\end{figure}

Approximately 95 per cent of our sample are low-excitation radio galaxies, with high-excitation radio galaxies making up the remaining $\sim$5 per cent, consistent with typical RLAGN distributions in the local Universe. Fig. \ref{fig:frac_lerg_herg} summarizes this distribution across radio luminosity, where we observe a decline in the LERG fraction and a corresponding rise in the HERG fraction around $10^{26}~\mathrm{W~Hz^{-1}}$, representing the well-known radio luminosity break between these two populations \citep{1996AJ....112....9L}. We discuss the morphological classifications within each emission-line sample in relation to the analysis presented in previous sections, focusing on the relationship between emission-line and radio properties.

\subsection{The nature of FRIs} \label{sec:fr1_sources}

\begin{figure*}
    \centering
    \raisebox{-0.1cm}{\includegraphics[width=\columnwidth]{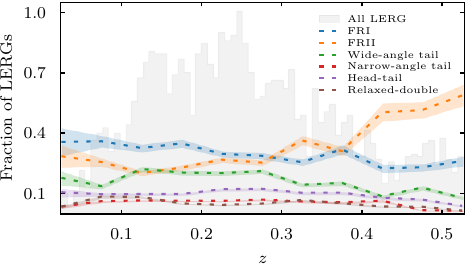}} \hspace{5mm}
\raisebox{-0.2cm}{\includegraphics[width=\columnwidth]{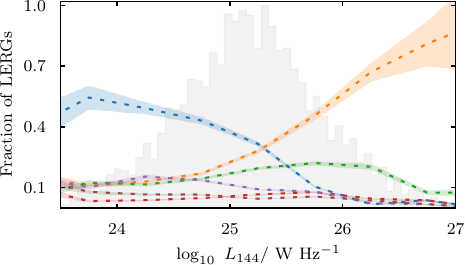}}
    \caption{Fraction of morphological classes in the LERG sample as a function of redshift (left panel) and radio luminosity (right panel). Shaded regions are depicted as in Fig. \ref{fig:frac_lerg_herg}.}
    \label{fig:frac_lerg_samp}
\end{figure*}

In previous sections, we have shown that FRIs represent a population of RLAGN characterized by diffuse jet structures extending from several kpc to a few Mpc, lower radio luminosities, and moderate to slightly higher core prominences, as is well known. Below, we discuss whether these properties are related to their optical emission line classifications.

Our sample reveals that the vast majority of FRIs exhibit a LERG spectrum, which aligns with the typical characteristics observed in RLAGN samples. Only six FRIs are classified as HERGs, which is consistent with the generally lower numbers of high excitation observed in FRI radio galaxies. Given the limited number of FRI HERGs in our sample, we focus primarily on FRI LERGs, as the small sample size of FRI HERGs does not provide enough statistical power for robust conclusions. However, we will briefly comment on the FRI HERG sample later in this section.

The redshift distribution of FRI LERGs in our sample is nearly uniform across the entire range considered ($z<0.57$), which is typical for LERGs at these low redshifts, as shown on the left side of Fig. \ref{fig:frac_lerg_samp}. This suggests that LERG morphological classifications in our sample can occur throughout the entire redshift range, with no clear morphological preference in any specific range. At radio luminosities lower than $10^{26}~\rm{W~Hz^{-1}}$, FRIs dominate the fraction of sources in the LERG sample, with a notable decline in counts at higher luminosities (right panel of Fig. \ref{fig:frac_lerg_samp}). This pattern indicates that FRIs are the preferred morphology at lower luminosities in LERG samples, while FRIIs become more prominent at higher luminosities -- we will expand on this point in Sec. \ref{sec:fr2_sources}.

It is well established that low-luminosity radio sources typically do not exhibit significant recent star formation \citep[e.g.][and references therein]{2017A&A...598A..49C}, with studies showing that these sources tend to favour host galaxies that are redder and more massive compared to their higher-luminosity counterparts \citep[e.g.][]{2008A&A...489..989B, 2010MNRAS.406.1841H}. This trend suggests that low-luminosity sources are often hosted in older, quiescent galaxies, where star formation has largely ceased, and the galaxy is in a more evolved state. In contrast, higher-luminosity sources are often associated with galaxies that exhibit ongoing star formation, where a reservoir of cold gas likely fuels both the AGN and star formation. Using mid-infrared diagnostics (Sec. \ref{sec:wise}), we find that a large fraction of low-luminosity FRI LERGs in our sample are hosted by redder hosts, in agreement with the results reported for the FRI catalogue compiled using NVSS, FIRST, and SDSS observations \citep[FRI$CAT$;][]{2017A&A...598A..49C}. Conversely, high-luminosity FRI LERGs in our sample seem to occupy the star-forming region of the WISE colour-colour plot. However, we lack sufficient data to confirm the presence of active star formation in these objects.

\begin{figure}
    \centering
    \includegraphics[width=\columnwidth]{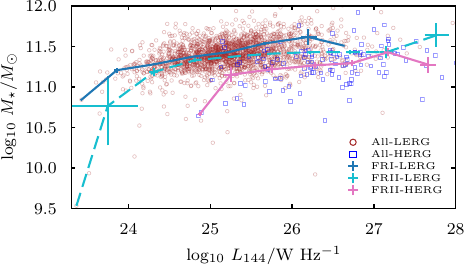}
    \caption{The dependence of radio luminosity on host galaxy mass, showing the distribution of FRI and FRII galaxies in the LERG/HERG sample. Errors and luminosity bin size as calculated in Fig. \ref{fig:lum_vs_R}.}
    \label{fig:mass_vs_lum_lerg}
\end{figure}

Although we lack specific environmental richness data to explore this trend further, we note a strong correlation between host galaxy mass and radio luminosity (Fig. \ref{fig:mass_vs_lum_lerg}). Less luminous FRI LERGs tend to reside in less massive hosts, while higher-luminosity FRI LERGs are found in more massive galaxies \citep[e.g.][]{2017A&A...598A..49C}. From this, we can infer that FRI LERGs at higher luminosities are more likely to inhabit richer environments compared to their lower-luminosity counterparts \citep[e.g.][]{2019A&A...622A..10C}. As a result, the host environment plays a crucial role in shaping the characteristics of the host galaxy \citep[e.g.][]{2013MNRAS.430.3086G} and the level of star formation rates surrounding the AGN, thereby influencing the nature of FRI sources.

\begin{figure}
    \centering
    \includegraphics[width=\columnwidth]{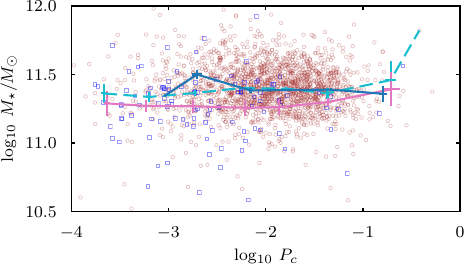}
    \caption{The dependence of core prominence on host galaxy mass for FRIs and FRIIs in the LERG/HERG sample. Error bars represent bootstrap estimates of bin mean values ($\Delta$log$_{10}~P_c = 0.5$) with 68 per cent confidence intervals. Legend same as Fig. \ref{fig:mass_vs_lum_lerg}.}
    \label{fig:mass_vs_R_lerg}
\end{figure}

In Fig. \ref{fig:mass_vs_R_lerg}, we examine the dependence of core prominence on host mass and observe no significant trend in the FRI LERG sample, indicating that stellar mass does not significantly impact core prominences, nor does it influence the formation of FRI jets. This outcome aligns with findings for more core-dominated sources, such as FR0s \citep[for recent reviews see][]{2023A&ARv..31....3B}
, which are also hosted by similar environments to those of FRIs. Thus, it is reasonable to conclude that core prominence has little to no bearing on whether an FRI source is classified as a HERG or LERG. Finally, we observe that the morphologies of FRI LERGs and FRI HERGs exhibit similar features, including the presence or absence of a radio core, a shared characteristic despite differences in optical emission line classification. Although limited by a small sample of FRI HERGs, these results suggest that the presence of a radio core is not solely determined by emission line classification.

\subsection{The nature of FRIIs}\label{sec:fr2_sources}
In recent years, a substantial number of FRII sources has been identified at lower flux limits through low-frequency surveys such as LoTSS, revealing a wide range of radio and emission-line properties. 

\citet{2019MNRAS.488.2701M} observed a substantial population of FRII sources spanning a broad range of radio luminosities, challenging traditional assumptions that high-luminosity FRIIs were the norm. This finding revealed a more diverse set of host galaxy environments and intrinsic galaxy properties within FRII populations. To explore this diversity, they divided the sample into low-luminosity and high-luminosity FRII sources, setting a threshold of approximately $10^{26}~\rm{W~Hz^{-1}}$ at 144 MHz to investigate differences between these populations. Their results showed that low-luminosity FRIIs, while traditionally thought to resemble FRI sources, still maintained the edge-brightened jet structures characteristic of FRIIs, indicating that morphology was not strictly tied to luminosity. Our sample shows similar trends, with a notably higher fraction of FRIIs at lower luminosities (69 per cent) compared to the 51 per cent of low-luminosity FRIIs reported by \citet{2019MNRAS.488.2701M}. FRIIs also dominate at the high-luminosity end (Sec. \ref{sec:pd}), as is well known. Given the substantial numbers of FRII LERGs and HERGs in our sample (Table \ref{tab:final_class}), the general picture is that low-luminosity FRIIs are primarily dominated by LERG-type spectra, whereas high-luminosity FRIIs are predominantly HERGs. This result aligns well with the findings of e.g. \citet{2017A&A...601A..81C}, who, in their compilation of FRII sources (FRII$CAT$) using NVSS, FIRST, and SDSS data, report a sample comprising 90 per cent LERGs and 10 per cent HERGs.

\begin{figure}
    \centering
    \includegraphics[width=\columnwidth]{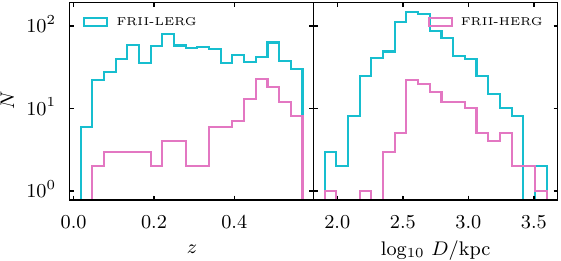}
    \caption{Comparison of HERG (pink) and LERG (cyan) distributions in redshift and physical size within the FRII Sample. The number of sources in this plot is given in the legend of Fig. \ref{fig:jet_powers}. Note that selection effects may influence the distribution of sources on these plots.}
    \label{fig:fr2_comp}
\end{figure}

From a morphological perspective, our inspection of FRII LERGs and FRII HERGs reveals no clear distinction between LERG and HERG classifications or between low- and high-luminosity sources. Both LOFAR and VLASS images show prominent features typical of FRII sources across the sample (Fig. \ref{fig:images} and \ref{fig:images2}), such as bright hotspots and/or radio cores, which are expected characteristics for the FRII population as a whole. Additionally, the redshift and size distributions for FRII HERGs and LERGs, as well as for high- and low-luminosity sources, show no significant differences from each other as shown in Fig. \ref{fig:fr2_comp}. 

\citet{2022MNRAS.511.3250M} also found that in the LoTSS Deep Fields \citep{2021A&A...648A...4D,2021A&A...648A...3K,2021A&A...648A...2S,2021A&A...648A...1T}, low-luminosity FRII sources are predominantly associated with LERGs, while high-luminosity FRIIs tend to be HERGs. This distinction suggests that while FRII morphology can emerge from both high- and low-efficiency accretion modes, the accretion mode significantly impacts the luminosity and emission-line properties of these sources. Our findings support this result, indicating that FRII morphologies may arise under a broader range of conditions than previously thought, with radio luminosity and accretion mode acting as distinct yet influential factors shaping the observable characteristics of radio galaxies. 

In Fig. \ref{fig:mass_vs_lum_lerg}, we observe a trend where radio luminosity increases with host mass across all FRII classifications. At a given host mass, FRII HERGs consistently exhibit higher radio luminosities than FRII LERGs. Conversely, at a given radio luminosity, FRII LERGs tend to have more massive hosts than FRII HERGs, supporting the idea that larger host galaxies can sustain more active AGN \citep[e.g.][]{2019A&A...622A..17S}. When comparing FRII LERGs with FRI LERGs, a distinction emerges between the two. Although this result may be influenced by selection effects, both classes exhibit a strong dependence of radio luminosity on host mass, particularly at the high-mass end. However, FRII LERGs appear to be more commonly hosted by a broader range of galaxies, though with a lower median mass compared to FRI LERGs, as previously noted \citep[e.g.][Sec. \ref{sec:host_mass}]{2022MNRAS.511.3250M}. 

We demonstrated in Sec. \ref{sec:core_pro} that FRII HERGs, on average, exhibit lower core prominences than FRII LERGs. However, further analysis of the relationship between core prominence and host galaxy mass (Fig. \ref{fig:mass_vs_R_lerg}) shows no dependence of core prominence on host galaxy mass in our sample, consistent with findings from the FRI sample in the previous section. Additionally, we previously highlighted that the differences in core prominence distributions between FRII LERGs and their HERG counterparts challenge the notion that FRII LERGs are simply `switched-off' FRII HERGs. This suggests intrinsic differences between the two populations, particularly in their accretion mechanisms. To explore this further, we examine the distribution of Eddington-scaled accretion rates within the FRII sample below. 

Following \cite{2014MNRAS.440..269M}, Eddington-scaled accretion rates can be estimated using the relation:

\begin{equation} \label{eq:one}
    \dot{m}_{\mathrm{Edd}} = \frac{L_{\mathrm{bol, [OIII]}} + Q_{\mathrm{jet}}}{L_{\mathrm{Edd}}}.
\end{equation}

Here, $L_{\mathrm{bol, [OIII]}}$ is the bolometric radiative luminosity, which is related to the [O{\scriptsize{III}}] luminosity by $L_{\mathrm{bol, [OIII]}} = 3500L_{\rm{[OIII]}}$ \citep{2004ApJ...613..109H}. The Eddington luminosity is given by $L_{\mathrm{Edd}}=1.3 \times 10^{31}M_{\rm{BH}}/M_{\odot}\rm{W}$. To estimate the black hole mass $M_{\rm{BH}}$, we use the scaling relation $M_{\rm{BH}} \approx 0.0014 M_{\star}$, where the black hole mass is approximately 0.14 per cent of host galaxy mass \citep{2004ApJ...604L..89H}.

The mechanical luminosity of the jet (jet power, $Q_{\mathrm{jet}}$) can be estimated using the relation $Q_{\mathrm{jet}} \propto L_{\mathrm{radio}}^{\beta}$, where $\beta\sim0.6-0.9$ depending on the assumptions of the model \citep[e.g.][]{2010ApJ...720.1066C,2012MNRAS.423.2498D,2014ARA&A..52..589H,2017MNRAS.467.1586I} and $L_{\mathrm{radio}}$ is the radio luminosity. In this work, we adopt the approach described by \citet{2019A&A...622A..12H} in estimating $Q_{\mathrm{jet}}$, which does not rely on this scaling relation. We apply a dynamical model of RLAGN evolution that takes into account observable parameters such as redshift, source size, and 144 MHz radio luminosity. Although we lack other parameters such as environmental richness and angle to the line of sight for each of our sources to refine this model, \citet{2019A&A...622A..12H} 
 showed that the model provides a robust statistical estimate of $Q_{\mathrm{jet}}$, particularly for large samples, as it accounts for radiative losses, inverse-Compton effects, and general environmental properties (e.g., gas density and temperature in galaxy groups or clusters).  

\begin{figure}
     \centering
     \includegraphics[width=\columnwidth]{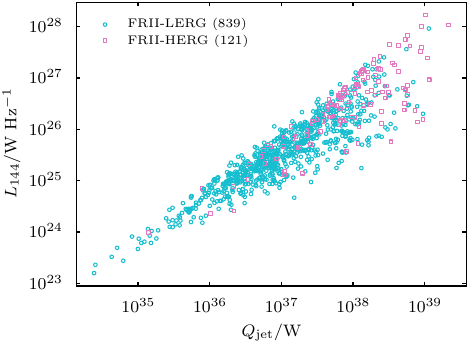}
     \caption{Distribution of the LERG and HERG populations in the FRII sample on the $Q_{\mathrm{jet}}-L_{\mathrm{144}}$ plane, with jet powers estimated using the methods of \citet{2019A&A...622A..12H}.}
     \label{fig:jet_powers}
 \end{figure}

\begin{figure}
    \centering
    \includegraphics[width=\columnwidth]{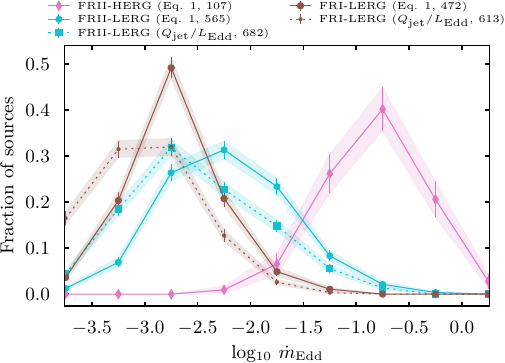}
    \caption{Eddington-scaled accretion rates for the FRII HERG and FRII LERG samples, with the FRI LERG sample included for comparison. Note the difference in the FRII populations in the legend of this plot and Fig. \ref{fig:jet_powers}; only sources with reliable host mass estimates (see Sec. \ref{sec:host_mass}) are shown here. Separate plots are shown for the LERG sample, distinguishing between sources with and without bolometric radiative luminosities estimated from [O{\scriptsize{III}}] measurements, as some LERGs lack these measurements. Shaded areas as in Fig. \ref{fig:frac_lerg_herg}.}
    \label{fig:fr2_comp2}
\end{figure}

 Fig. \ref{fig:jet_powers} shows the distribution of the LERG and HERG population within the FRII sample on the $Q_{\mathrm{jet}}-L_{\mathrm{144}}$ plane, calculated using the methods of \citet{2019A&A...622A..12H}. We observe a strong correlation between jet power and radio luminosity, as we would expect if jet powers were estimated using the jet power-radio luminosity scaling discussed above, and we use these results to estimate Eddington-scaled accretion rates. Fig. \ref{fig:fr2_comp2} shows the distribution of these Eddington-scaled accretion rates estimated using Eq. \ref{eq:one}. For sources lacking [O{\scriptsize{III}}] measurements, we approximate $\dot{m}_{\mathrm{Edd}} \approx Q_{\mathrm{jet}}/L_{\mathrm{Edd}}$, a particularly useful approach for LERGs, as they are inefficient at producing [O{\scriptsize{III}}] emission, which, when measured, is often boosted by the jet rather than accretion. Comparing the LERG sample under both methods of estimating $\dot{m}_{\mathrm{Edd}}$ -- with and without including radiative luminosity -- confirms that they display consistently low accretion rates, as expected \citep[e.g.][]{2014MNRAS.440..269M,2025MNRAS.536..554K}. Conversely, the HERG sample has reliable [O{\scriptsize{III}}] measurements, allowing their Eddington rates to be estimated solely using Eq. \ref{eq:one}. 

We see that our results agrees with previous studies \citep{2012MNRAS.421.1569B,2014MNRAS.440..269M} and theoretical models, which predict that the accretion mode in RLAGN transitions at a specific threshold of Eddington-scaled accretion rates, typically occurring around log$_{10}~\dot{m}_{\mathrm{Edd}} \sim$ $-$2 and $-$3, depending on factors like black hole spin \citep[e.g.][]{2012MNRAS.421.1569B, 2022MNRAS.516..245W} and surrounding environmental conditions. The FRII HERGs exhibit higher Eddington ratios (above one per cent), consistent with expectations for sources operating in RE accretion mode. In this regime, the accretion disk is optically thick and geometrically thin, emitting substantial thermal energy in the optical and UV bands. In contrast, FRII LERGs accrete below one per cent Eddington, with their distribution indicating a transition to RI accretion mode. This regime is characterized by a geometrically thick and optically thin accretion structure, such as an ADAF or RIAF, where most of the energy remains as kinetic output in jets rather than being radiated.

Although we observe some overlap between FRII HERGs and FRII LERGs, a clearer separation emerges below one per cent Eddington when comparing FRII HERGs with FRI LERGs. This distinction highlights the different accretion modes between FRI LERGs (and LERGs more broadly) and FRII HERGs. However, we note that the jet-launching mechanisms in FRI LERGs and FRII LERGs could share similarities. Based on core prominence and these results, it is reasonable to conclude that FRII LERGs represent a distinct population, exhibiting intermediate properties between FRI LERGs and FRII HERGs \citep[e.g.][Sec. \ref{sec:core_pro}]{2020MNRAS.493.4355M}. However, additional data, such as the particle content in their lobes \citep[e.g.][]{2008MNRAS.386.1709C,2018MNRAS.476.1614C}, is needed to draw robust conclusions.

\subsection{The nature of bent-tail and relaxed double sources}
\subsubsection{WATs, NATs, and HTs}
In this section, we discuss WATs, NATs, and HTs. These objects have been classified using various methods, with our work employing a visual morphological classification (Sec. \ref{sec:radio_morp}). While they are commonly considered FRIs, they appear to be more closely associated with galaxy clusters \citep[e.g.][]{2001AJ....121.2915B,2007ApJS..172..295S,2022AJ....163..280M,2023A&A...674A.191M} rather than the galaxy groups typically favoured by FRIs. Nonetheless, they represent an important subgroup of radio galaxies, essential for understanding the overall evolution of radio galaxies and their environments.

It is well known that narrow and wide angle tails are found in richer environments \citep[typically in clusters showing signs of merging rather than relaxed clusters; e.g.][]{2022AJ....163..280M} compared to the general RLAGN population at the same radio luminosity. WATs tend to favour the cores of galaxy clusters, where systematic gas motions are intense, while NATs are typically located away from the cluster centers, where gas motions are less pronounced. Head-tails, on the other hand, largely share the same environments as NATs, leading to suggestions that these objects are simply unresolved NATs observed face-on \citep[e.g.][]{2017A&A...608A..58T}. This environmental difference is thought to play a key role in shaping their structures. 

\begin{figure}
    \centering
    \includegraphics[width=\columnwidth]{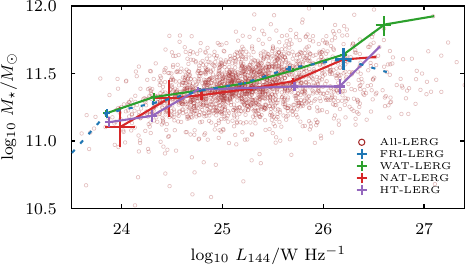}
    \caption{As in Fig. \ref{fig:mass_vs_lum_lerg}, but now showing the bent-tail LERG sources from our sample. The FRI plot is also included for reference, shown as dotted lines.}
    \label{fig:bent_sources_M_vs_L}
\end{figure}

We lack cluster information for these sources to further examine their hosts. However, using mid-IR data (Fig. \ref{fig:wise_cc2}), we observe no significant differences between the hosts of WATs, NATs, and HTs compared to the general RLAGN population in our sample, as noted by several authors previously \citep[e.g.][and references therein]{2019A&A...626A...8M,2023Galax..11...67O}. Despite this, the host masses differ, with WATs favouring more massive hosts than NATs and HTs, as shown in Fig. \ref{fig:steler_mas} and Fig. \ref{fig:bent_sources_M_vs_L}. In the radio luminosity versus host galaxy mass plane (Fig. \ref{fig:bent_sources_M_vs_L}) and in terms of physical extent (Fig. \ref{fig:pd_diagram}), these sources overlap with FRI sources, including their core prominences (Sec. \ref{sec:core_pro}). This highlights the idea that bent sources are simply highly distorted FRI sources, with their morphologies influenced by ram pressure as they move through galaxy clusters of different gas densities and motions. The fact that a large fraction of these sources are classified as LERGs with only a small minority spectroscopically classified as HERGs (Table \ref{tab:final_class}) further supports this suggestion.

\subsubsection{RDs}
Relaxed doubles represent a unique population of radio galaxies that could provide valuable insights into the life cycles of AGN activity.

Approximately 5 per cent of sources in our sample were classified as RDs, with four identified as HERGs and the remaining 149 as LERGs (Table \ref{tab:final_class}). Although we do not see notable differences between RDs and the general population in our sample in terms of radio power–linear size relationship (Fig. \ref{fig:pd_diagram}), mid-IR diagnostics (Fig. \ref{fig:wise_cc2}), host galaxy mass (Fig. \ref{fig:steler_mas}), or core prominence distributions (Fig. \ref{fig:core_prom}), these sources exhibit interesting morphological features (Fig. \ref{fig:images} and \ref{fig:images2}), such as amorphous lobes that resemble those seen in FRII sources. Examination of RD HERGs and RD LERGs provides no clear indication of a relationship between emission-line class (though the HERG sample is small) and morphology. However, we observe core detections at higher frequencies in some objects from both groups, suggesting one of two possibilities: either a restarting activity phase or a fading radio core before transitioning into the remnant phase. We discuss this further in the next section (\ref{sec:other_classes}).

\subsection{Subcategories: Remnant, Restarted, and Giant candidates}\label{sec:other_classes}
Here, we discuss the construction of a subsample of remnant and restarted RLAGN candidates within our sample. The section progresses by discussing radio and emission-line properties of remnant and restarted candidates (Sec. \ref{sec:remt_rest}) and we examine a population of giant RLAGN candidates in Sec. \ref{sec:giants}.

\subsubsection{Remnant and Restarted candidates}\label{sec:remt_rest}
In Sec. \ref{sec:core_pro} we showed that the analysis of core prominences ($P_c$) in our sample agrees with trends reported in the literature but also raises questions on the validity of using $P_c$ to distinguish between active, remnant, and restarted sources. We apply a visual classification to select remnant and restarted candidates, to investigate how well $P_c$ would classify our objects as remnants.

We classified our remnant candidate sample by first selecting all sources without core detections in VLASS images (3 GHz central frequency), resulting in 501 sources. As noted by e.g. \citet{2018MNRAS.475.4557M}, a non-detection at higher frequencies does not necessarily indicate a remnant phase, and we found similar results (see Fig. \ref{fig:images} and \ref{fig:images2} for images of sources that are without core detections at 3 GHz yet appear active at 144 MHz). Many of the 501 sources show activity at 144 MHz -- such as bright radio lobes or cores -- so we excluded these. The remaining 70 sources meet the criteria for remnant phase classification, exhibiting amorphous structures with no defined lobes or cores at 144 MHz.

Methods for selecting restarted candidates in the literature are somewhat ambiguous due to limited surface brightness sensitivity in most surveys, which often leads to misclassification. For example, hotspots in WATs can be mistaken for double lobes if the diffuse emission from WATs is undetected, and jet knots in FRIs can also resemble double lobes. To clarify distinctions between active and restarted sources, we retained only FRII and RD classifications. Within these, we selected only FRIIs that resemble DDRGs \citep[e.g.][]{2000MNRAS.315..371S,2019A&A...622A..13M,2024arXiv240813607D}, resulting in 48 sources. For RDs, we included only sources with core detections at both 144 MHz and 3 GHz, requiring steep core spectral indices \citep[$\alpha>0.5$; e.g.][]{1996MNRAS.278....1A} and amorphous structures (low surface brightness lobes). This resulted in a sample of 19 RDs that are potential restarted candidates. Our final sample therefore comprises a total of 67 restarted candidates.

\begin{table}
\centering
    \caption{Morphological breakdown of selected remnant and restarted candidates based on visual inspection, with the number of HERGs in each category indicated in brackets.}
    \begin{tabular}{c|c|c}
     \hline\hline
     Morphology & Remnant & Restarted \\
     & candidates & candidates \\
     \hline
     FRI  & 8 [1] &   \\
     FRII & 20 & 48 [6] \\
     WAT  & 1 &   \\
     NAT  & 1 &   \\
     HT   & 1 &   \\
     RD   & 39 [1] & 19 [2]\\
     \hline
     Total  & 70 [2] & 67 [8] \\
     \hline\hline
    \end{tabular}
    \label{tab:other_classes}
\end{table}

\begin{figure}
    \centering
    \includegraphics[width=\columnwidth]{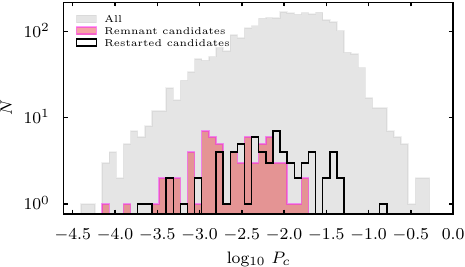}
    \caption{Core prominences for candidate remnant and restarted sources in our sample, with the general RLAGN population shown in the background.}
    \label{fig:remnt_rest_R}
\end{figure}

Table \ref{tab:other_classes} summarizes the classifications of remnant and restarted candidates along with their emission-line distributions. Fig. \ref{fig:remnt_rest_R} displays the distribution of core prominences for the final sample of these candidates. As previously noted, the continuous distribution of core prominences in Fig. \ref{fig:core_prom} does not serve as a reliable selection criterion for these sources, and the final selection in Fig. \ref{fig:remnt_rest_R} confirms this observation. While a significant fraction of candidate remnants fall below the proposed threshold of log$_{10}~P_c<-2$, this pattern does not apply to restarted candidates, with only a small proportion above the suggested threshold of log$_{10}~P_c>-1$. Consequently, a single $P_c$ value cannot be used to select these sources in our case study. Nonetheless, their morphologies suggest that these sources could indeed be potential remnant and restarted candidates.

\begin{figure}
    \centering
    \includegraphics[width=\columnwidth]{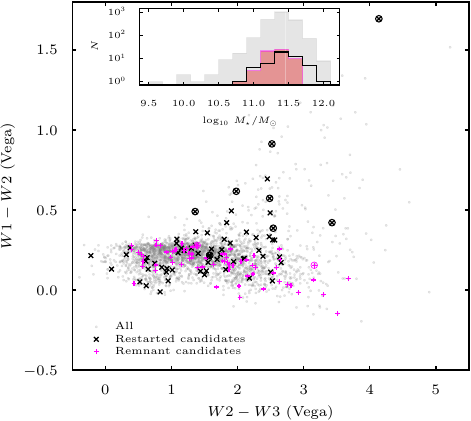}
    \caption{Mid-IR environments of remnant and restart candidates. The inset shows the distribution of host galaxy mass for these categories, overlaid on the general RLAGN population in our sample. Circles highlight, respectively, remnant and restart candidates spectroscopically classified as HERGs.}
    \label{fig:rem_rest_hosts}
    
\end{figure}

Fig. \ref{fig:rem_rest_hosts} shows the host environments of remnant and restarted candidates based on mid-IR information and host galaxy masses. The findings reveal that both types of candidates span a range of environments -- from red, massive elliptical galaxies to bluer, star-forming galaxies. This overlap suggests a short interval between the remnant and restarted stages, enabling both populations to appear in similar host environments \citep[e.g.][and references therein]{2023Galax..11...74M}. This pattern supports the hypothesis of a rapid transition between these phases. Notably, we observe a few restarted candidates, including some restarted HERGs, within the AGN-dominated region, whereas remnant candidates are absent from this area, as expected.

Our study of remnant and restarted candidates aligns well with previous studies \citep[e.g.][]{2017A&A...606A..98B, 2018MNRAS.475.4557M, 2019A&A...622A..13M,2021MNRAS.508.5326M,2021A&A...653A.110J}, though here we can draw more robust conclusions due to having optical emission-line classifications for all sources in our sample. Most remnants and restarted candidates in our sample are classified as LERGs, although we do observe a few remnant HERGs, which is unexpected since these sources are thought to be in a fading stage, generally lacking the nuclear activity that produces strong ionization. Examining their morphologies confirms they are likely true remnant candidates. A closer look at their emission-line classifications reveals that these HERGs were classified based on the specific [O{\scriptsize{III}}] approach shown in Fig. \ref{fig:o3m_lherg}, falling below the 80 per cent reliability threshold considered in this work \citep[classified as HERG with scores around 50 per cent and 75 per cent for the two sources;][]{2024MNRAS.534.1107D}. However, Fig. \ref{fig:o3m_lherg} shows overlapping distributions, highlighting the classification’s limitations, as expected for a method that is not perfectly accurate.

We also examined the morphologies of restarted candidates classified as HERG and LERG, finding no morphological distinctions between the two, as highlighted in earlier sections for other morphological classifications. These candidates display a broad range of sizes, with around 37 per cent having sizes over 700 kpc. Notably, five out of the eight HERG restarted sources fall into this large-size category, indicating that restarted HERGs can reach extensive sizes similar to their LERG counterparts. This suggests that emission-line classification alone may not adequately predict the AGN duty cycle, as both HERG and LERG restarted sources share overlapping characteristics in scale and structure. 

In conclusion, our findings imply that factors beyond emission-line class -- such as environmental influences or episodic accretion \citep[see e.g.][]{2020NewAR..8801539H} -- may play significant roles in the AGN duty cycle.

\subsubsection{Giant candidates}\label{sec:giants}
We identified a subset of RLAGN whose radio emission extends beyond 700 kpc, commonly referred to as giant radio galaxies \citep[GRGs; e.g.][]{1974Natur.250..625W,1999MNRAS.309..100I,2000MNRAS.315..371S,2020A&A...635A...5D,2020A&A...642A.153D,2021MNRAS.501.3833D,2022A&A...660A...2O,2022MNRAS.515.2032S,2024Natur.633..537O,2024ApJS..273...30B}. The structures of GRGs include radio lobes, jets, and sometimes hotspots, extending far beyond their host galaxies (see images in Fig. \ref{fig:images} and \ref{fig:images2}). Although GRGs are considered rare compared to `normal' radio galaxies (RGs), the primary distinction observed so far is their age, as there is no evidence suggesting they constitute a unique population of AGN \citep[e.g.][]{2021MNRAS.502.5104L}. 

\begin{figure*}
    \includegraphics[width=\columnwidth]{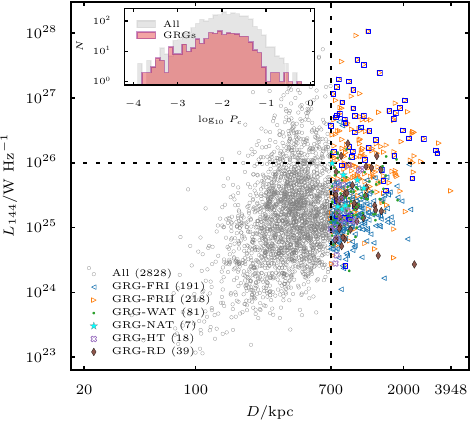}\hspace{5mm}
    \includegraphics[width=\columnwidth]{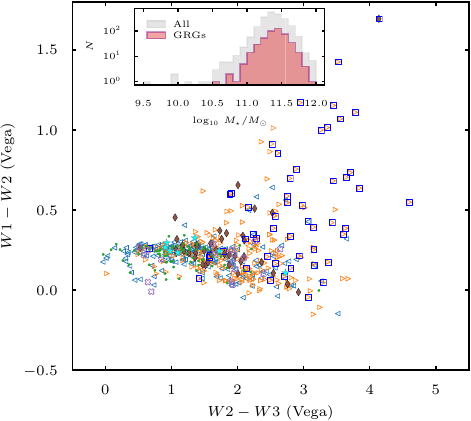}
    \caption{Left: P-D diagram of GRG candidates with core prominence histogram insert. The vertical line marks the GRG size threshold (above 700 kpc), and the horizontal line separates low- and high-luminosity GRGs at around $10^{26}~\rm{W~Hz^{-1}}$. Right: GRG environments in WISE colour-colour space with host mass histogram insert. HERGs are marked by squares.}
    \label{fig:giants}
\end{figure*}

Previous studies indicate that GRGs predominantly exhibit FRII morphologies, while FRI giants are less common due to their diffuse radio emissions, which are often undetected by most radio surveys \citep[e.g.][]{2021MNRAS.501.3833D}. However, in our sample, we observe that although FRII giant candidates dominate (218 out of 554), a significant fraction also displays FRI morphologies (191 out of 554). Additionally, our sample includes giant candidates with WAT (81 out of 554), NAT (7 out of 554), HT (18 out of 554), and RD (39 out of 554) morphologies, suggesting that GRGs can exhibit various forms. Further information, such as cluster data for these bent-tailed GRGs, however, is needed to confirm whether these galaxies are indeed bent-tailed GRGs, although it has been established that a small population of GRGs resides in cluster environments \citep[e.g.][]{2020A&A...642A.153D}. Therefore, it is not surprising that we observe a significant population of bent-tailed GRGs, given that WATs and NATs, for instance, tend to favour such environments \citep[e.g.][]{2004MNRAS.349..560H}.

Although the largest GRG reported to date extends up to 7 Mpc \citep{2024Natur.633..537O}, GRGs with sizes greater than 2 Mpc are rare in GRG catalogues, comprising only about 9 per cent \citep{2020A&A...642A.153D}. We observe a similar trend in our sample, with only 4 per cent (22 out of 554) of GRGs exceeding 2 Mpc. The largest object in our sample has a physical size of approximately 3.9 Mpc and is classified as an FRII. This source has a radio luminosity of around $3 \times 10^{25}\ \rm{W~Hz^{-1}}$ and is classified as a LERG, consistent with findings for most of the largest GRGs \citep[e.g.][]{2020A&A...642A.153D,2021MNRAS.501.3833D}.

The P-D space of Fig. \ref{fig:giants} highlights a population of HERG GRGs (52). Most of these are classified as FRIIs (47), with a smaller number being FRIs (3), WATs (1), and RDs (1). This observation supports the idea that the physical size of GRGs is not determined solely by their accretion mode but mainly determined by the age of the radio source. While HERG GRGs are present, indeed, GRGs are predominantly LERGs \citep[e.g.][]{1998MNRAS.298.1035T}, as evidenced by the large number of LERG GRGs in our sample (502). These LERGs are morphologically diverse, including 188 FRIs and 171 FRIIs. Other morphological classifications, such as WATs (80), NATs (7), HTs (18), and RDs (38), are also represented. Notably, there is no significant difference in the host environments of GRGs compared to the overall RLAGN sample, as demonstrated by the WISE colour-colour space and host galaxy mass in Fig. \ref{fig:giants}. Additionally, GRGs exhibit a wide range of core prominences, consistent with the presence of restarted and remnant sources, which represent older populations of RLAGN.

\section{Conclusions}\label{sec:concl}
We have visually constructed a sample of 2828 extended, nearby ($z < 0.57$) radio-loud active galactic nuclei based on the LOw-Frequency Array Two-metre Sky Survey second data release \citep{2022A&A...659A...1S,2023A&A...678A.151H}. The sample is classified into various morphological categories, including Fanaroff \& Riley type I and type II sources, wide-angle and narrow-angle tail sources, head-tail sources, and relaxed double sources. Selection criteria were based on flux completeness ($>$10 mJy) and extended emission ($>$60 arcsec). 

Optical emission-line classifications \citep{2024MNRAS.534.1107D} were also used, enabling us to differentiate between high-excitation and low-excitation radio galaxies. Below, we present a summary of our results, which aim to explore the relationship between optical emission-line and radio properties for a large, representative sample:

\begin{itemize}
    \item [(i)] We used the power–linear size (P–D) diagram to examine the distribution of our visually classified morphologies and found that most objects occupy their expected regions. However, we also identify a minority of non-FRII objects (e.g. RDs) in the high-luminosity region, where they would not typically be expected. We find that a large fraction of optically classified sources in the high-luminosity region are HERGs, mostly of FRII morphology, while LERGs dominate the low-luminosity region of the P-D diagram, with both FRIs and FRIIs present. As expected, almost all FRI sources in our sample exhibit a LERG spectrum.

    \item [(ii)] A probe into their environments using mid-infrared diagnostics reveals that many of our LERGs (including large fractions of FRI, FRII, WAT, NAT, HT, and RD sources) are associated with redder, older elliptical galaxies. In contrast, the HERGs (mostly FRII sources) tend to reside in bluer, less massive, star-forming galaxies. This is consistent with their host galaxy mass distributions, which span moderate to massive galaxies, with some FRII LERGs exhibiting a tail of lower masses, as reported in previous studies.   

    \item [(iii)] We analysed VLASS core prominences to assess radio core brightness relative to extended emission across both morphological and emission-line classifications, finding a wide range of values typical for RLAGN. Our findings indicate that FRIs are more core-dominant than FRIIs. As expected, WATs, NATs, and HTs exhibit intermediate core prominences between FRI and FRII sources, likely due to their gas- and dust-rich environments, which can cause rapid jet deceleration and thus higher core prominences than FRIIs (especially HERGs), which generally inhabit sparser environments. Notably, FRII LERGs tend to have higher core prominences than their HERG counterparts. Overall, we conclude that HERGs in our sample show lower core prominences on average than LERGs.

    \item [(iv)] We observe that about 99 per cent of the FRI sample is classified as LERGs, with the majority of these sources exhibiting low luminosities. The FRII sample is also dominated by LERG sources, many of which are low-luminosity objects. However, we observe that the HERG sample is predominantly composed of FRII sources, allowing us to draw significant conclusions about their Eddington-scaled accretion rates. We show that FRII LERGs span a wide range of luminosities, which evolve with the masses of their host galaxies. Similar trends are observed in the FRI LERG sample, although FRI LERGs appear to favour more massive galaxies at the same radio luminosity as FRII LERGs. The FRII LERG sample is more biased towards higher luminosities, with luminosity steeply increasing as host galaxy mass increases. However, we find no significant dependence of core prominences on host galaxy mass. This leads us to conclude that core prominences do not have a significant impact on emission-line classifications.

    \item [(v)] We also find that bent-tail sources (WATs, NATs, and HTs) as well as relaxed doubles are predominantly LERGs, sharing similar radio and emission-line properties with FRIs. In contrast, RDs exhibit properties similar to FRIIs, despite lacking the prominent signatures of AGN activity typically seen in FRIIs. Although a few HERGs are present in these classifications, their morphologies do not exhibit structures unique to the LERG population. Further information about the environments of these sources is needed to better understand the accretion modes in these samples.  

    \item [(vi)] We show that while some remnant and restarted candidate RLAGN may exhibit lower and higher core prominences, respectively, compared to the general RLAGN population, using core prominence as a criterion for identifying these sources may risk overlooking significant remnant and restarted populations. Nonetheless, core prominences provide valuable insight into the accretion modes of this subclass of sources.

    \item [(vii)] Lastly, we identify a substantial population of RLAGN with radio emission extending beyond 700 kpc, classifying them as giant radio galaxies, and spanning a wide range of morphologies. Notably, both HERGs and LERGs are present within this category of RLAGN, with HERGs comprising approximately 38 per cent of the population. This shows the diverse excitation mechanisms operating in GRGs, despite their exceptional physical sizes.
\end{itemize}

In our upcoming paper in the series, we aim to investigate the relationship between core spectral indices and core prominence in this sample, to better understand the physical processes occurring in the central regions of these galaxies. It will also be useful to investigate the spectral ages of this sample to gain insights into the AGN duty cycle, which will require multi-frequency observations at comparable resolutions, from e.g. the LOFAR Low Band Antenna Sky Survey \citep[LoLSS;][]{2021A&A...648A.104D,2023A&A...673A.165D}. Additionally, a deeper understanding of the relationship between morphology and emission-line classification in these radio-loud AGN would require examining their environmental richness \citep[e.g.][]{2019A&A...622A..10C} and particle content \citep[e.g.][]{2018MNRAS.476.1614C}. Together, these analyses will shed light on the external and internal factors shaping the observed properties of these sources, enhancing our understanding of AGN evolution and host interactions.

\section*{Acknowledgements}
JC acknowledges financial support from the University of Hertfordshire through a PhD studentship. MJH and JCSP acknowledge financial support from STFC [ST/V000624/1, ST/Y001249/1].

LOFAR is the Low Frequency Array, designed and constructed by ASTRON. It has observing, data processing, and data storage facilities in several countries, which are owned by various parties (each with their own funding sources), and which are collectively operated by the ILT foundation under a joint scientific policy. The ILT resources have benefited from the following recent major funding sources: CNRS-INSU, Observatoire de Paris and Universit\'{e} d'Orl\'{e}ans, France; BMBF, MIWF-NRW, MPG, Germany; Science Foundation Ireland (SFI), Department of Business, Enterprise and Innovation (DBEI), Ireland; NWO, The Netherlands; The Science and Technology Facilities Council, UK; Ministry of Science and Higher Education, Poland; The Istituto Nazionale di Astrofisica (INAF), Italy.

This research made use of the Dutch national e-infrastructure with support of the SURF Cooperative (e-infra 180169) and the LOFAR e-infra group. The J\"{u}lich LOFAR Long Term Archive and the German LOFAR network are both coordinated and operated by the J\"{u}lich Supercomputing Centre (JSC), and computing resources on the supercomputer JUWELS at JSC were provided by the Gauss Centre for Supercomputing e.V. (grant CHTB00) through the John von Neumann Institute for Computing (NIC).

This research made use of the University of Hertfordshire high performance computing facility and the LOFAR-UK computing facility located at the University of Hertfordshire (\url{https://uhhpc.herts.ac.uk}) and supported by STFC [ST/P000096/1], and of the Italian LOFAR IT computing infrastructure supported and operated by INAF, and by the Physics Department of Turin University (under an agreement with Consorzio Interuniversitario per la Fisica Spaziale) at the C3S Supercomputing Centre, Italy.

This research made use of \texttt{ASTROPY}, a community-developed
core Python package for astronomy \citep{2013A&A...558A..33A},  of \texttt{APLPY}, an open-source astronomical plotting package for Python hosted at \url{http://aplpy.github.com/}, and of \texttt{TOPCAT} \citep{2005ASPC..347...29T}.

The National Radio Astronomy Observatory is a facility of the National Science Foundation operated under cooperative agreement by Associated Universities, Inc.

\section*{Data Availability}
The data used in this work is part of a LoTSS DR2 sample with emission-line classifications \citep{2024MNRAS.534.1107D} which can be accessed from \url{www.lofar-surveys.org}. The classifications used in this work can be shared upon reasonable request to the corresponding author.

\bibliographystyle{mnras}

\bibliography{References} 

\appendix 

\section{Images}
A representative sample of RLAGN based on visual classification is shown here. These include FRI, FRII, WAT, NAT, HT, and RD sources. Additionally, we showcase a subset of remnant, restarted, and giant candidate RLAGN.

\begin{figure*}
    \includegraphics[width=0.65\columnwidth]{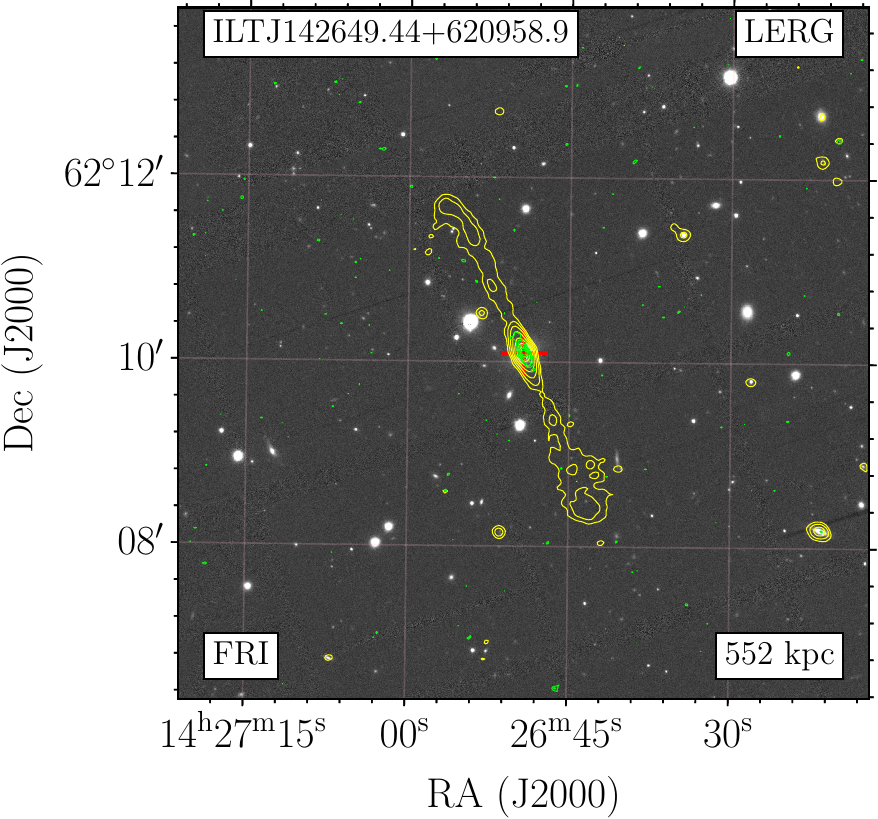}
    \includegraphics[width=0.65\columnwidth]{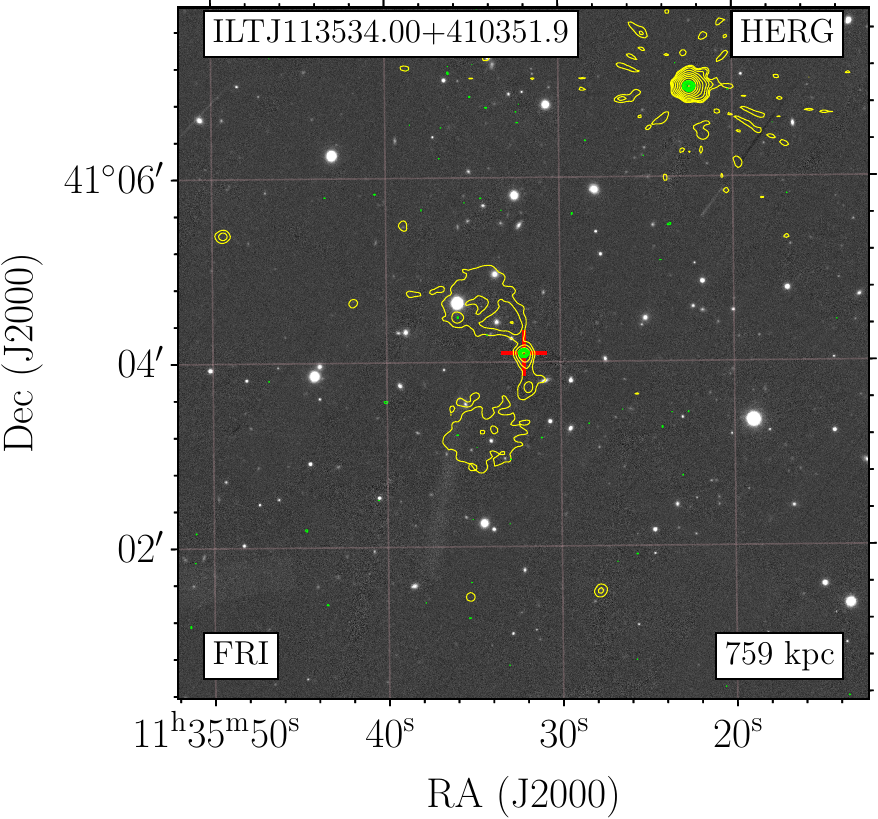}
    \includegraphics[width=0.65\columnwidth]{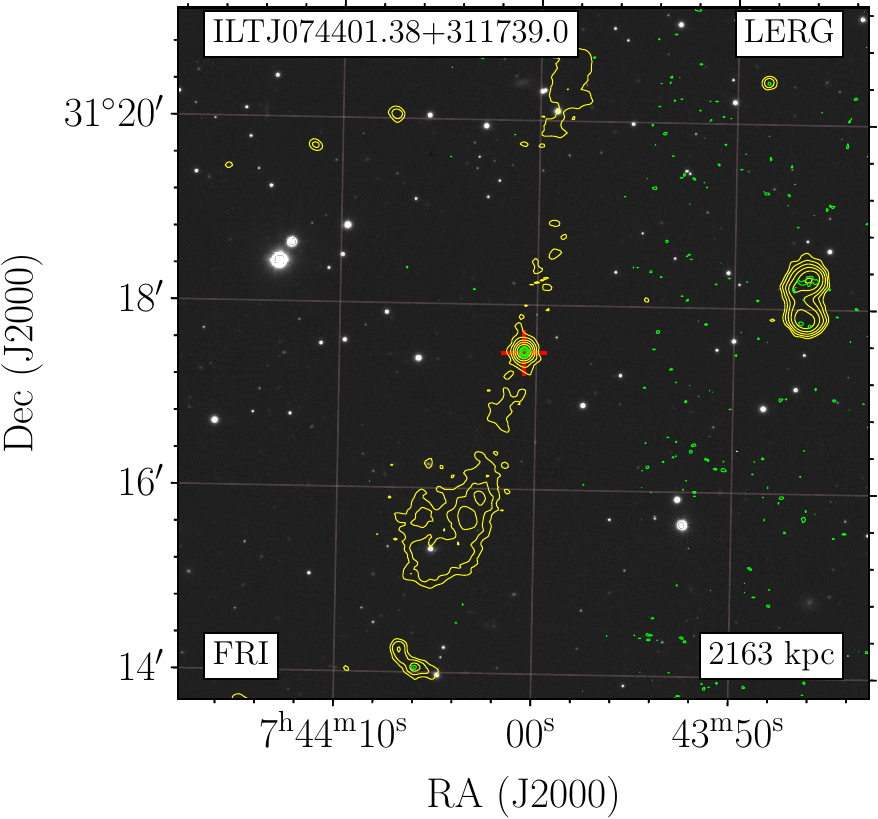}

    \includegraphics[width=0.65\columnwidth]{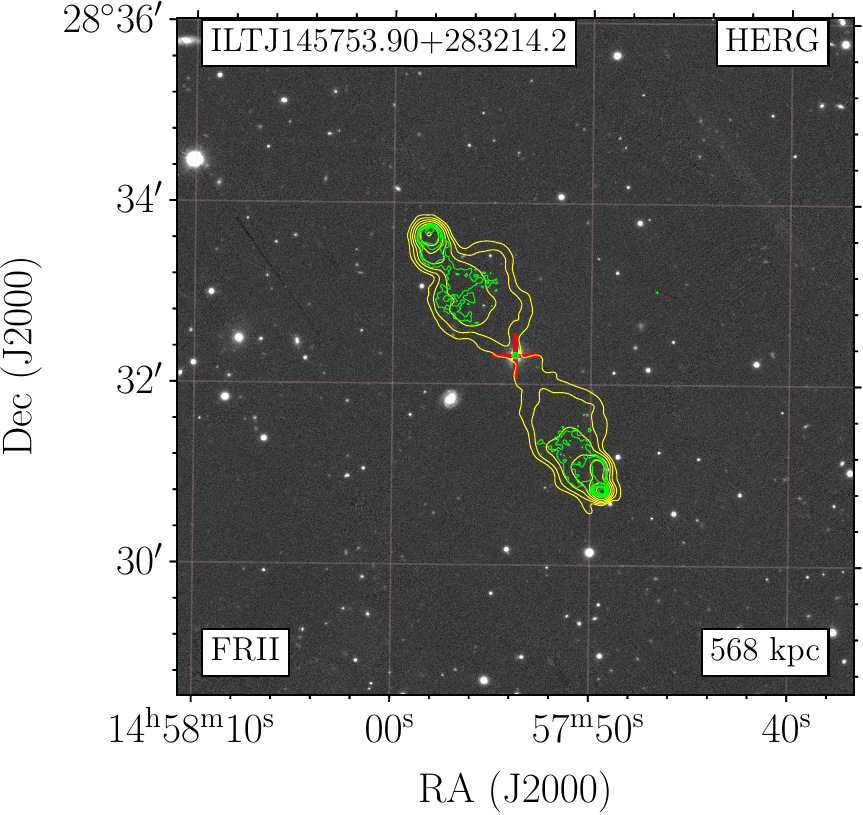}
    \includegraphics[width=0.65\columnwidth]{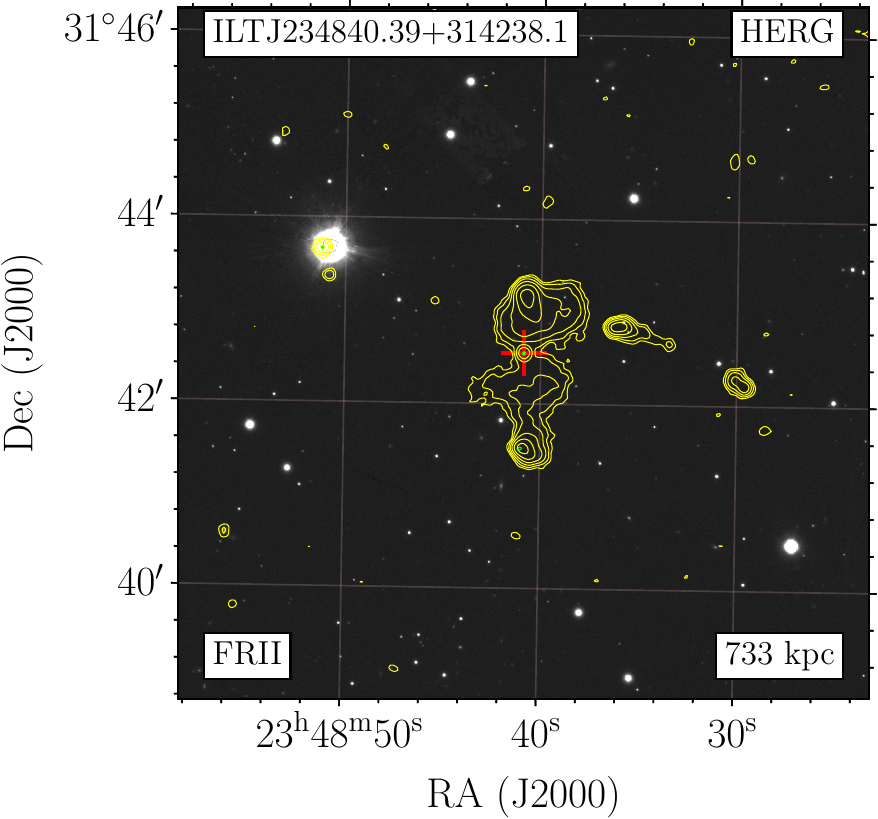}
    \includegraphics[width=0.65\columnwidth]{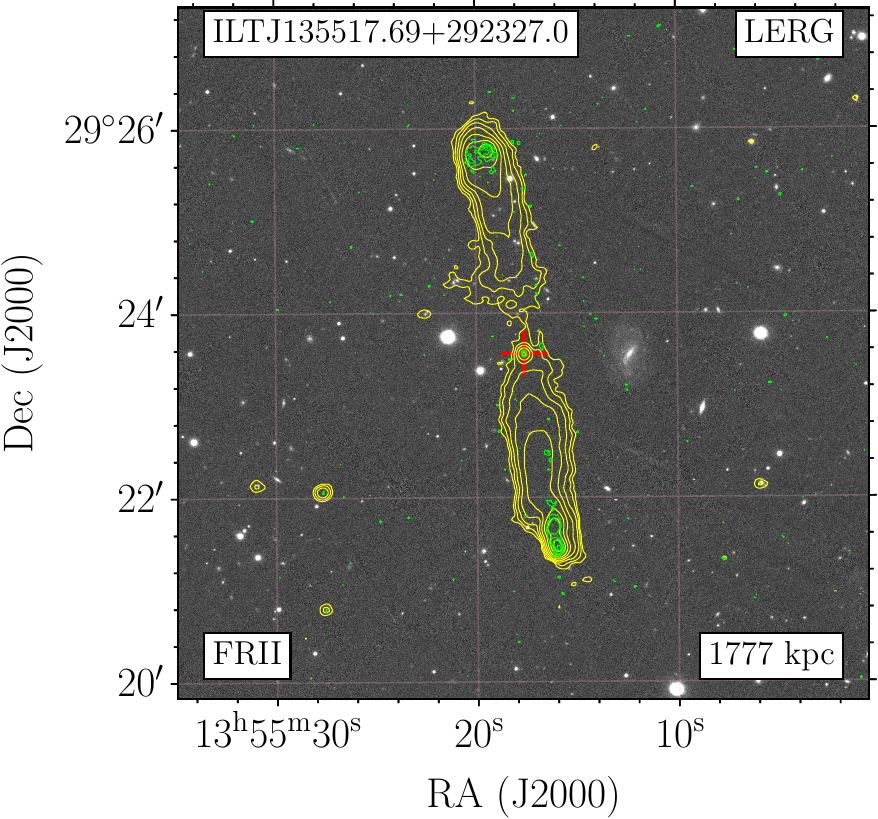}

    \includegraphics[width=0.65\columnwidth]{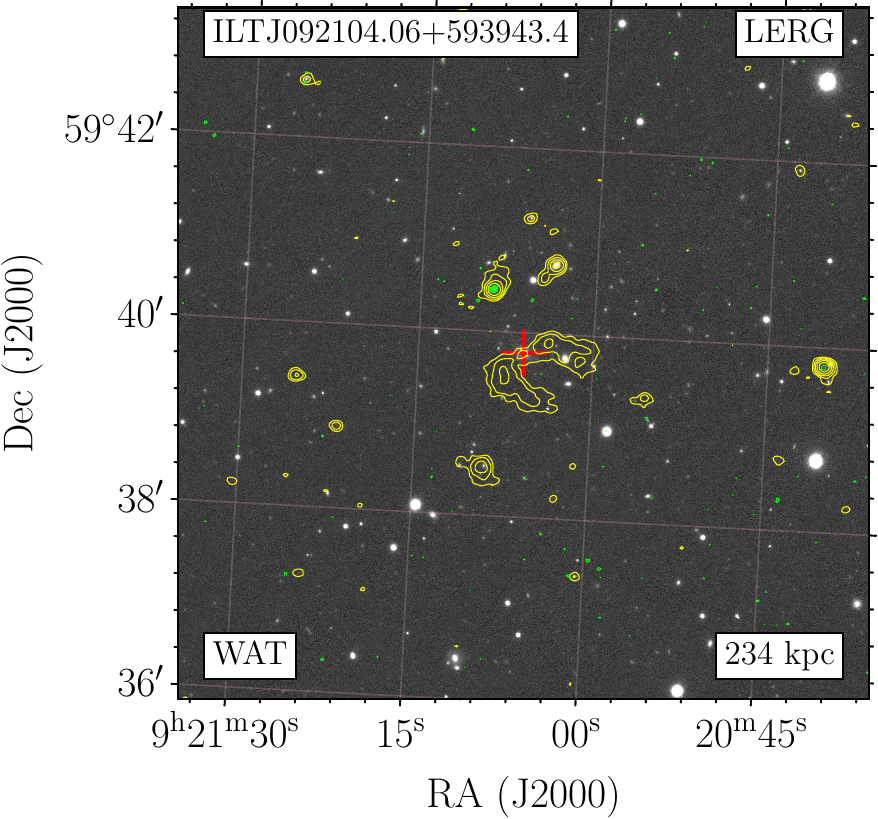}
    \includegraphics[width=0.65\columnwidth]{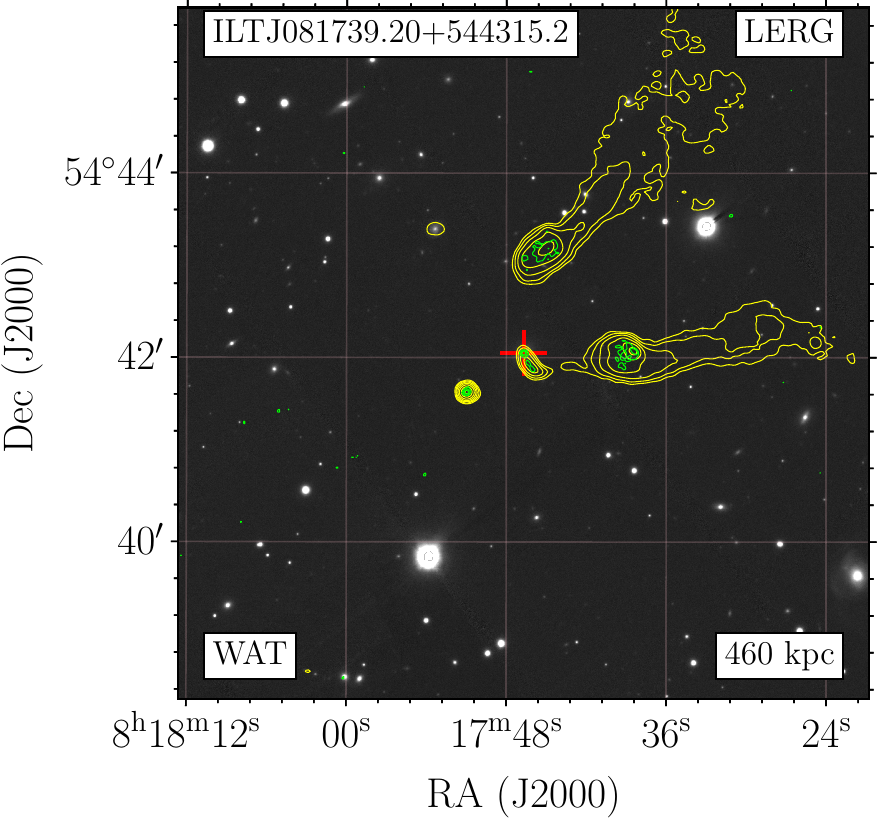}
    \includegraphics[width=0.65\columnwidth]{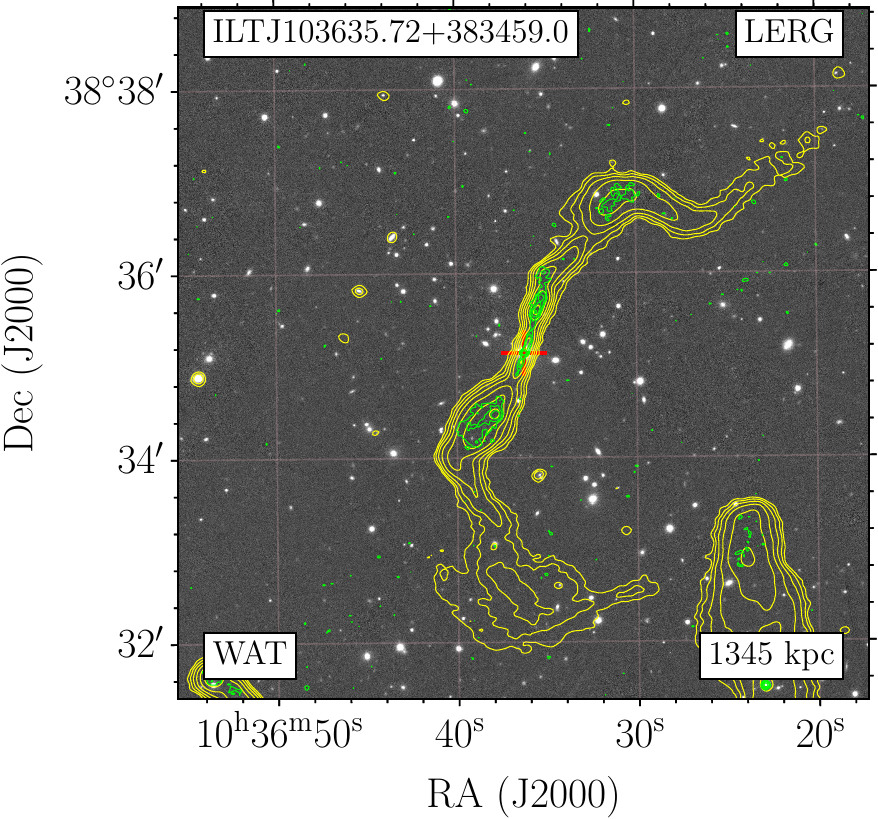}

     \includegraphics[width=0.65\columnwidth]{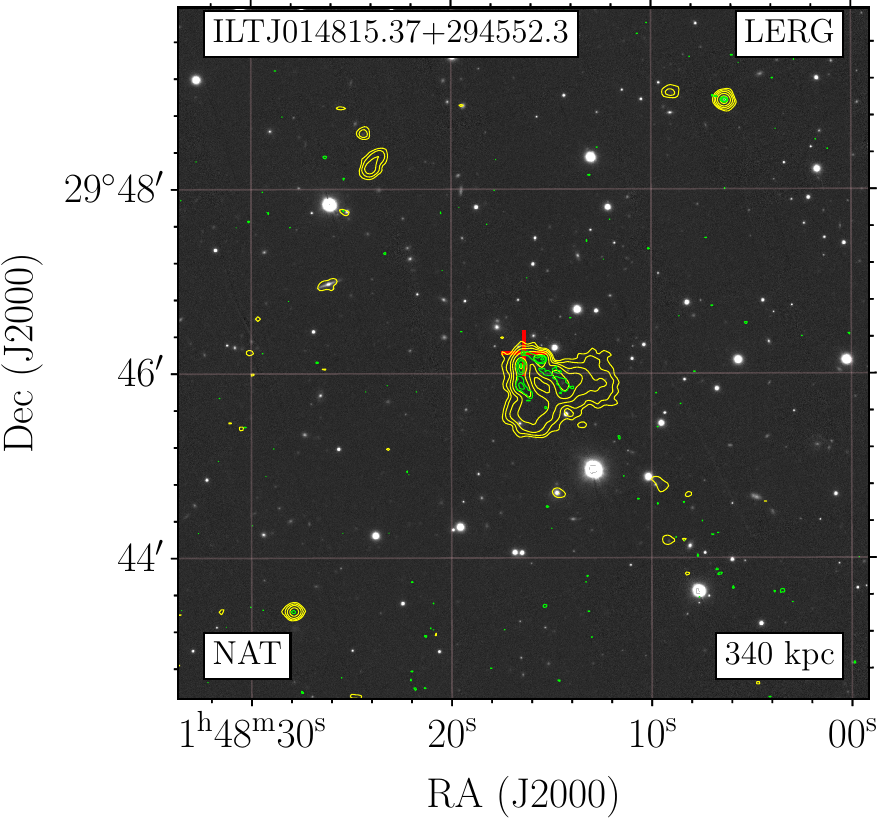}
    \includegraphics[width=0.65\columnwidth]{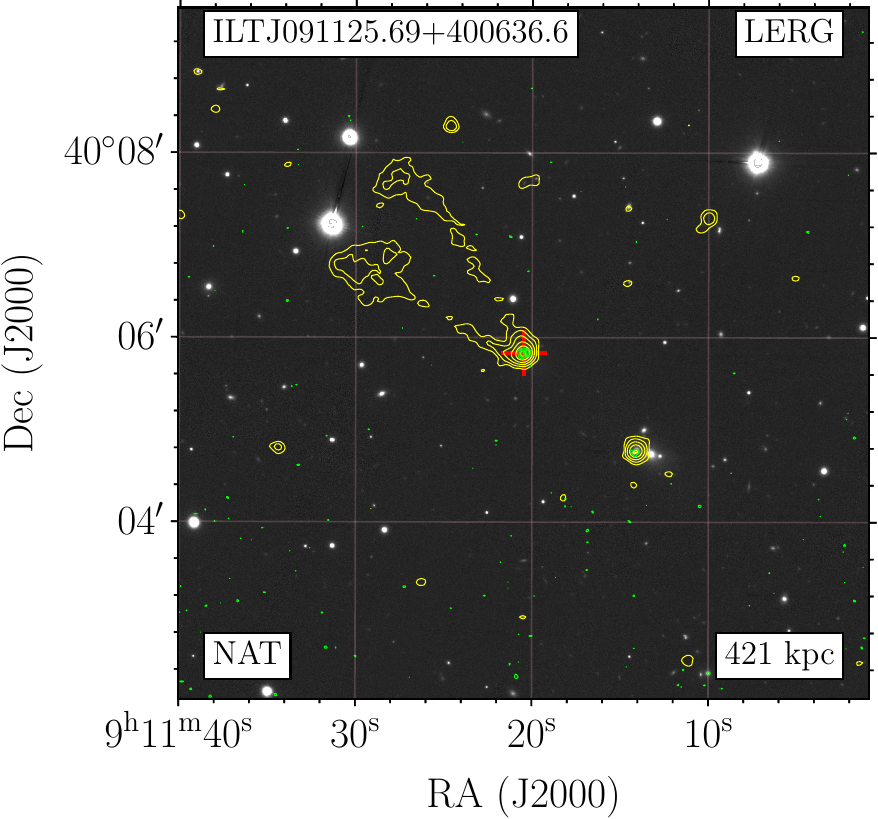}
    \includegraphics[width=0.65\columnwidth]{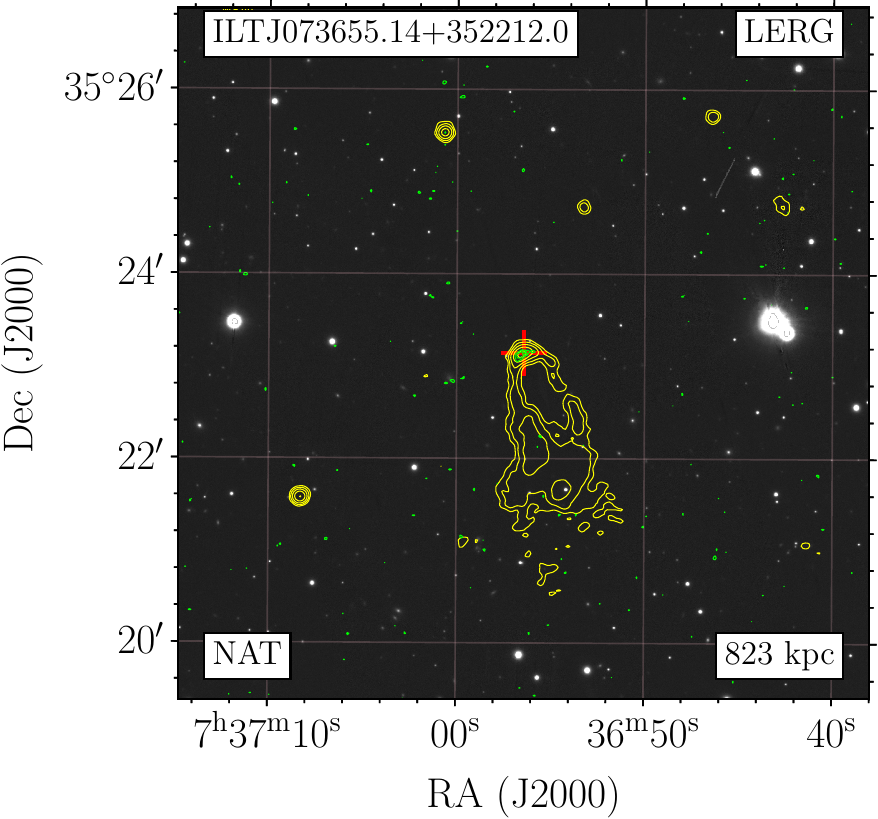}
    
    \caption{Representative images from our final morphological classifications are shown in this figure following visual inspection. LOFAR and VLASS contours, in gold and green respectively, are overlaid on i-band Pan-STARRS images, with optical IDs marked by a plus sign. Physical size estimates are provided in the bottom-right corner of each panel to highlight `normal' and candidate giant radio galaxies. Continued in Fig. \ref{fig:images2}.}
    \label{fig:images}
\end{figure*}

\begin{figure*}
    \includegraphics[width=0.65\columnwidth]{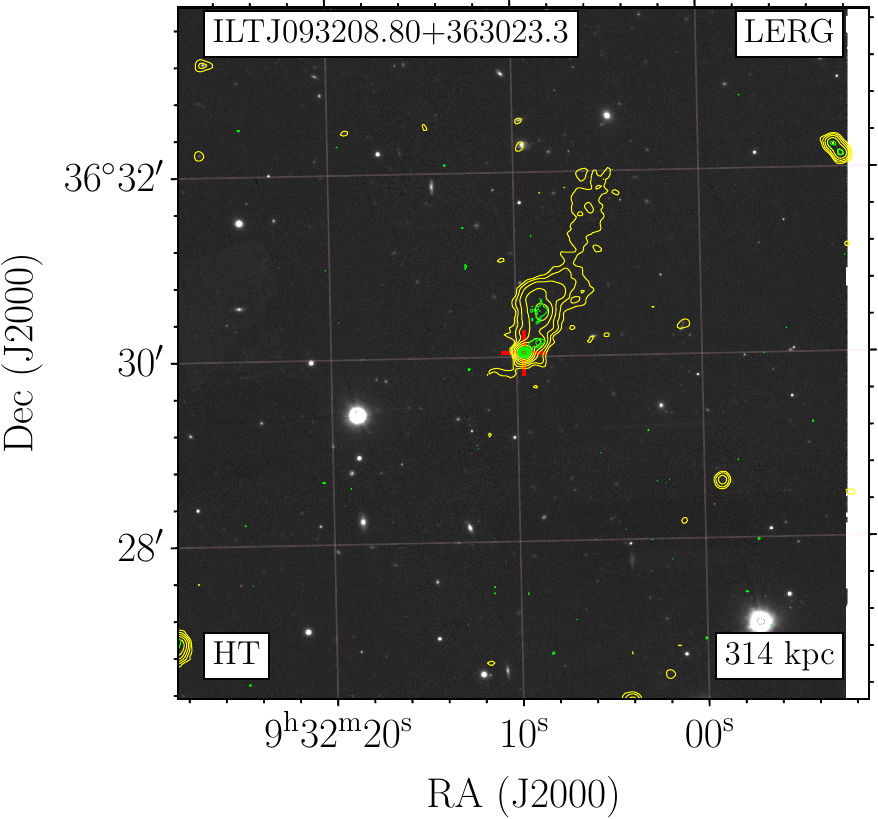}
    \includegraphics[width=0.65\columnwidth]{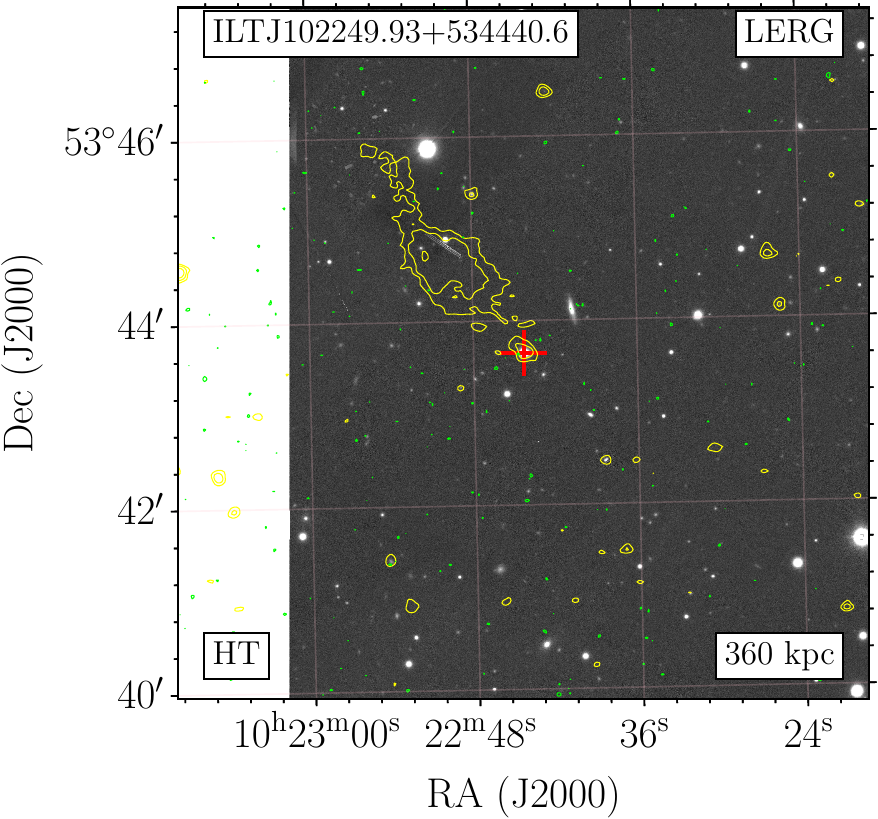}
    \includegraphics[width=0.65\columnwidth]{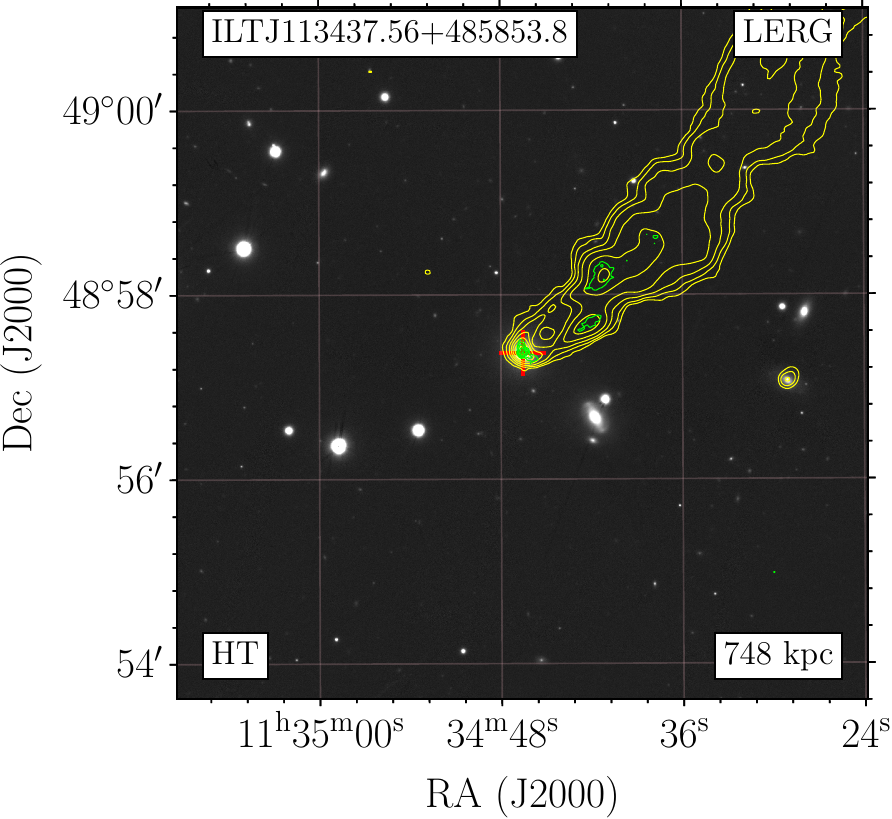}

    \includegraphics[width=0.65\columnwidth]{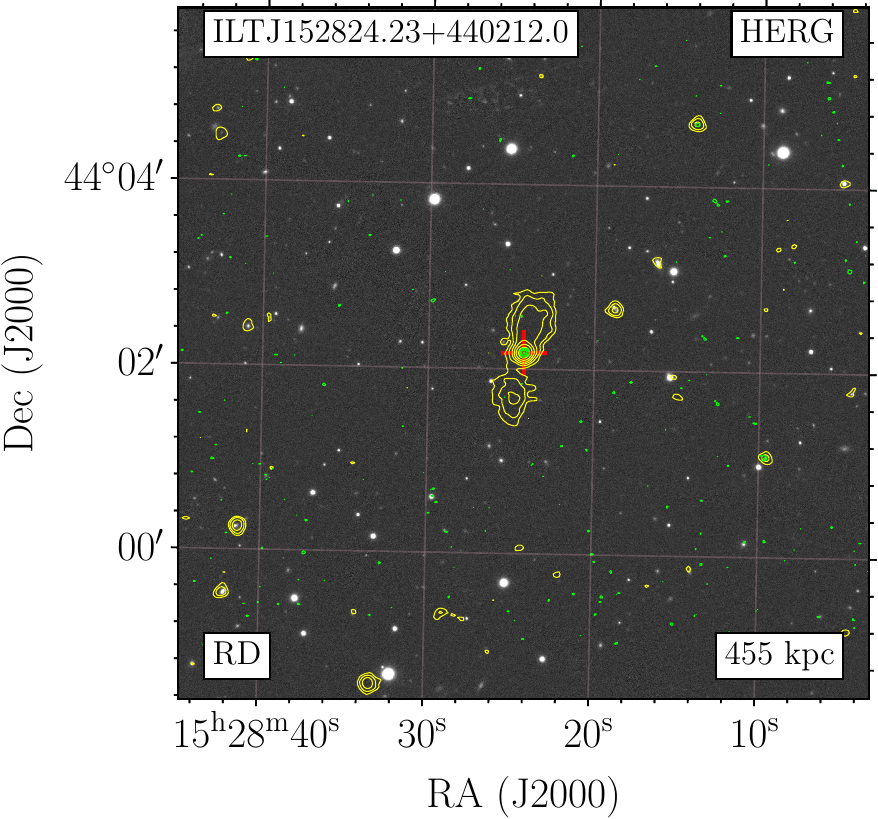}
    \includegraphics[width=0.65\columnwidth]{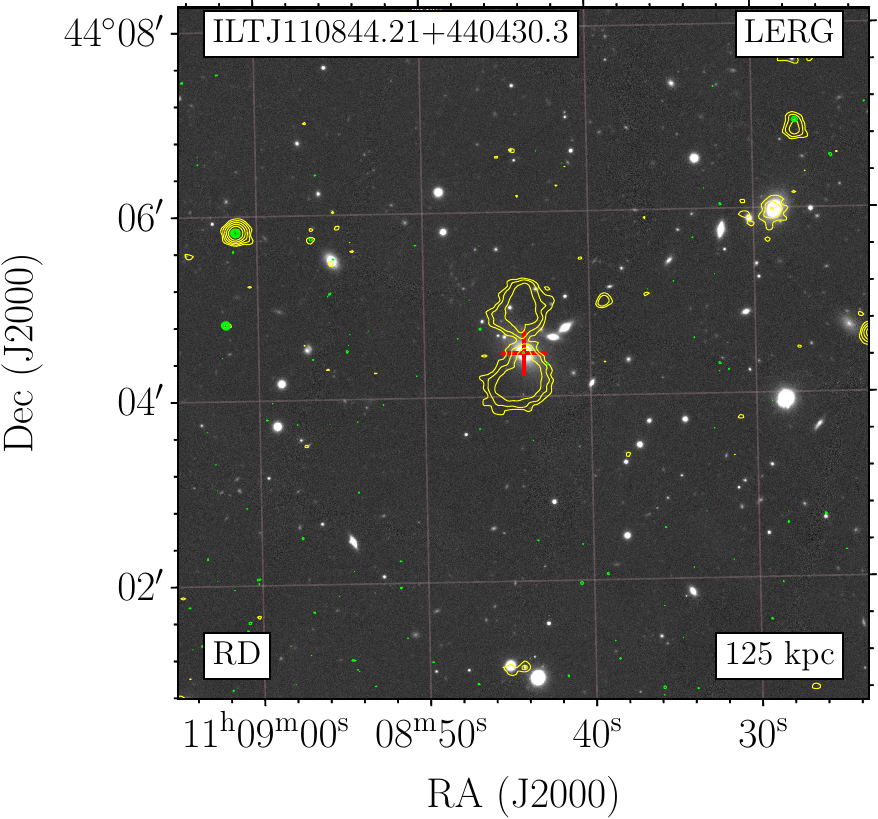}
    \includegraphics[width=0.65\columnwidth]{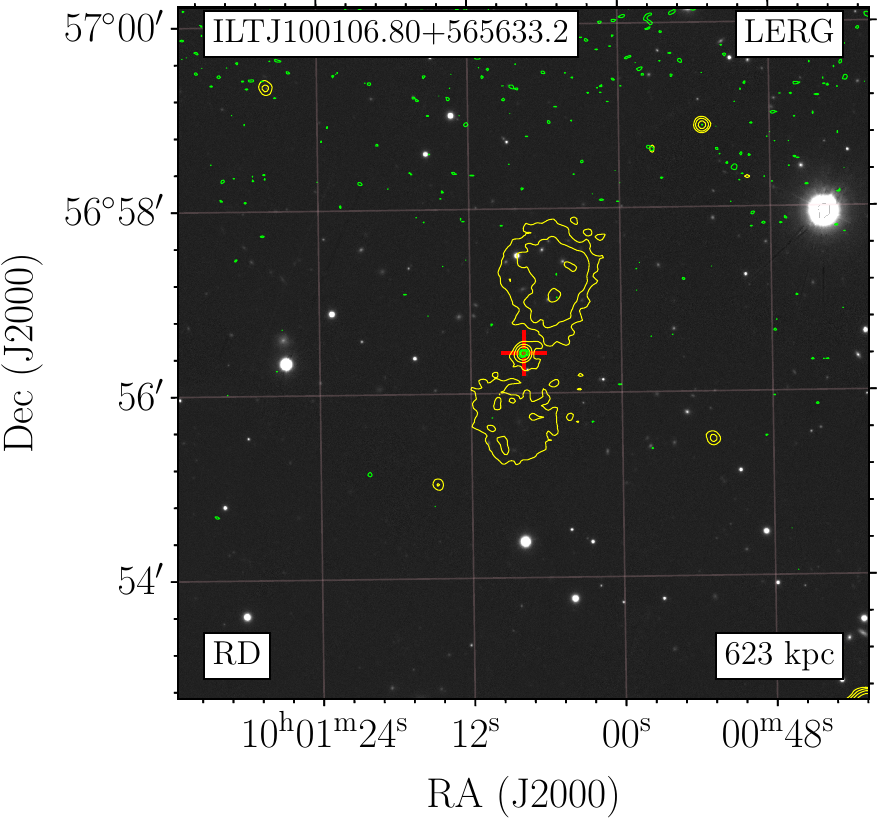}

    \includegraphics[width=0.65\columnwidth]{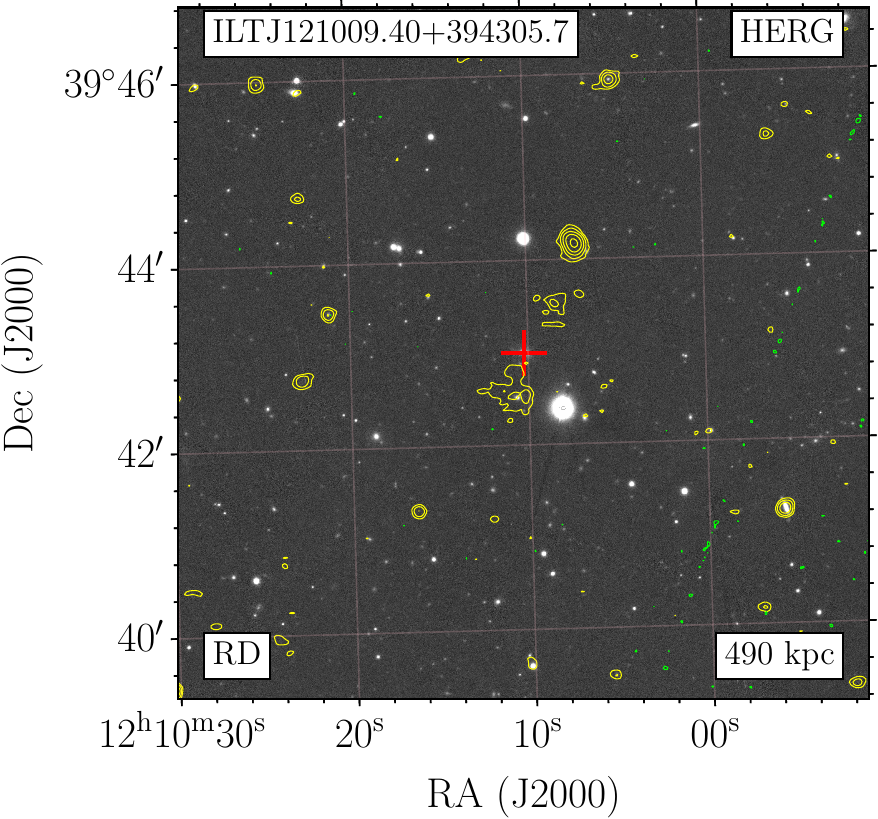}
    \includegraphics[width=0.65\columnwidth]{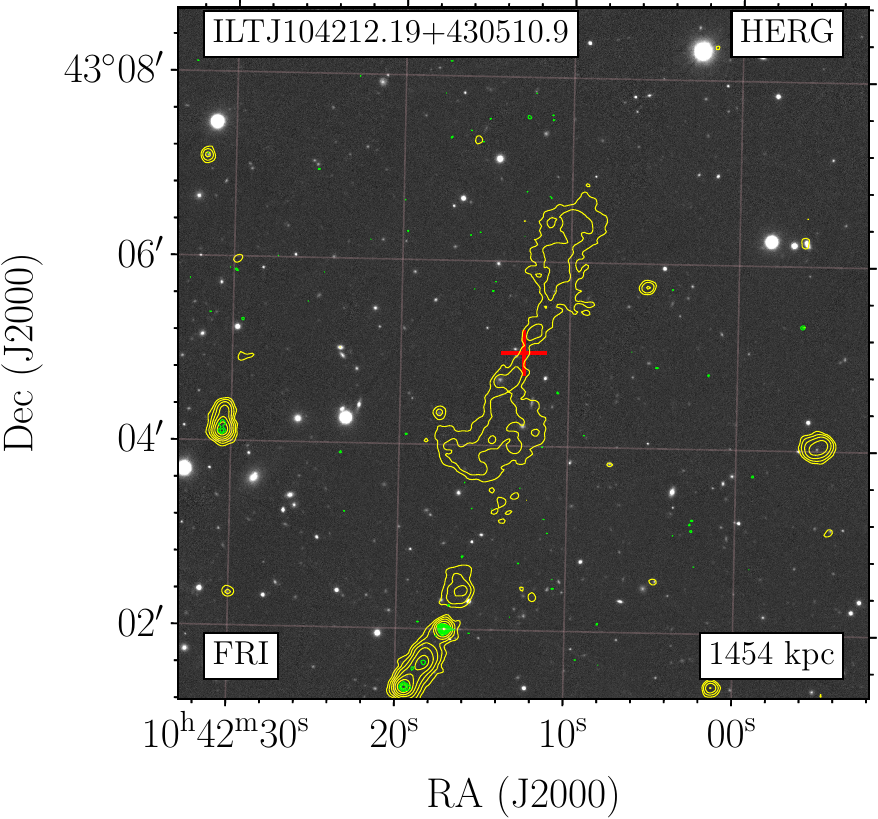}
    \includegraphics[width=0.65\columnwidth]{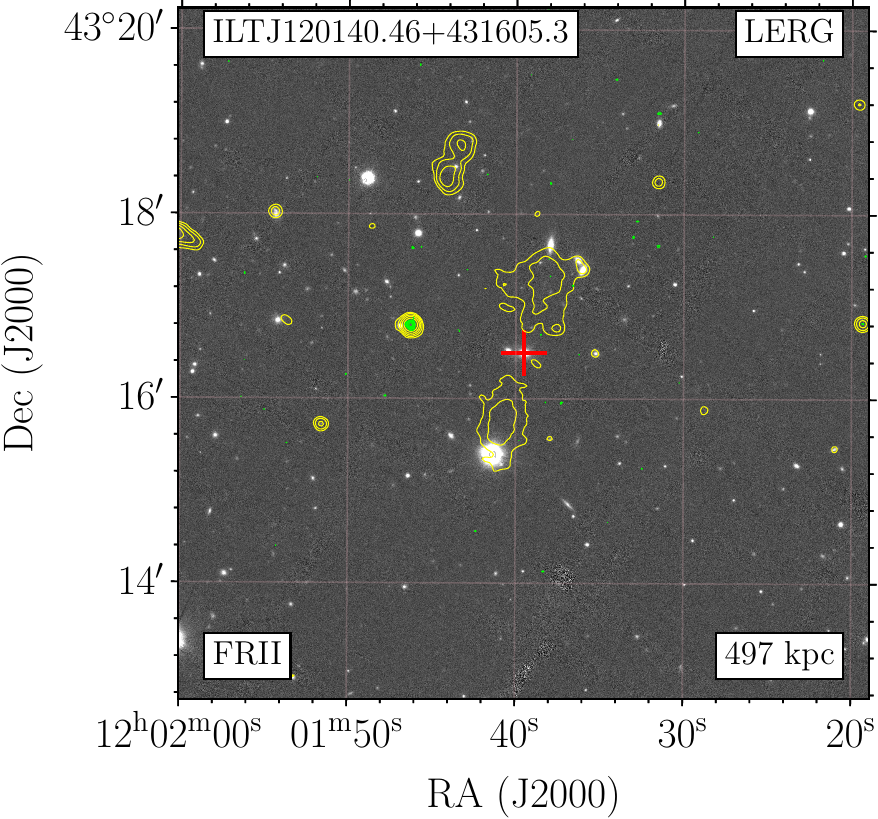}

    \includegraphics[width=0.65\columnwidth]{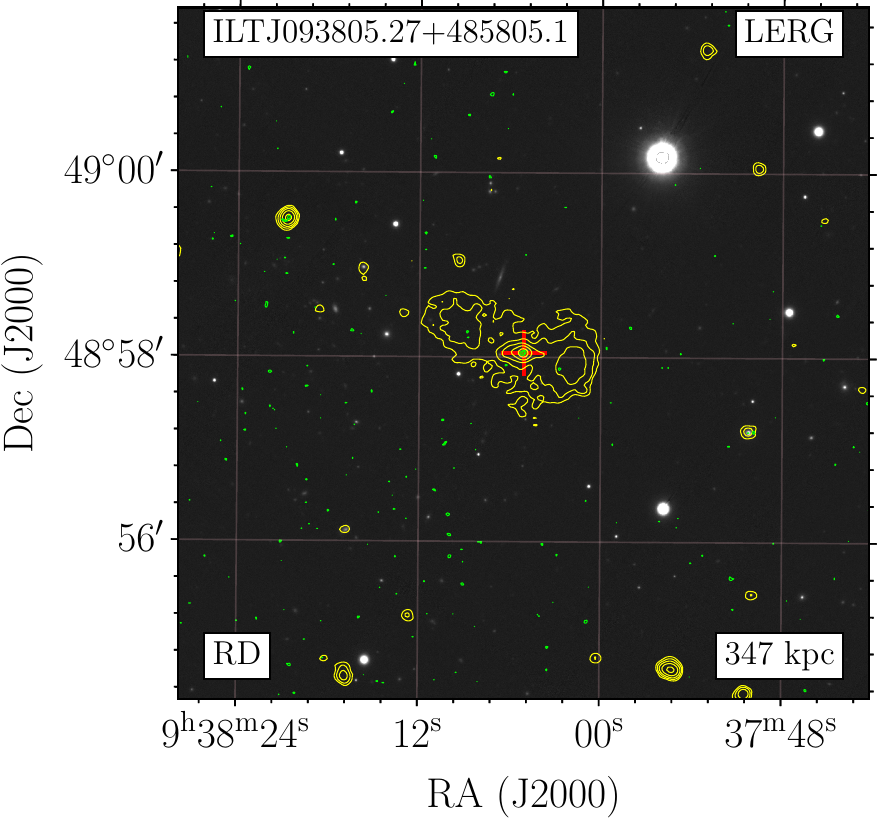}
    \includegraphics[width=0.65\columnwidth]{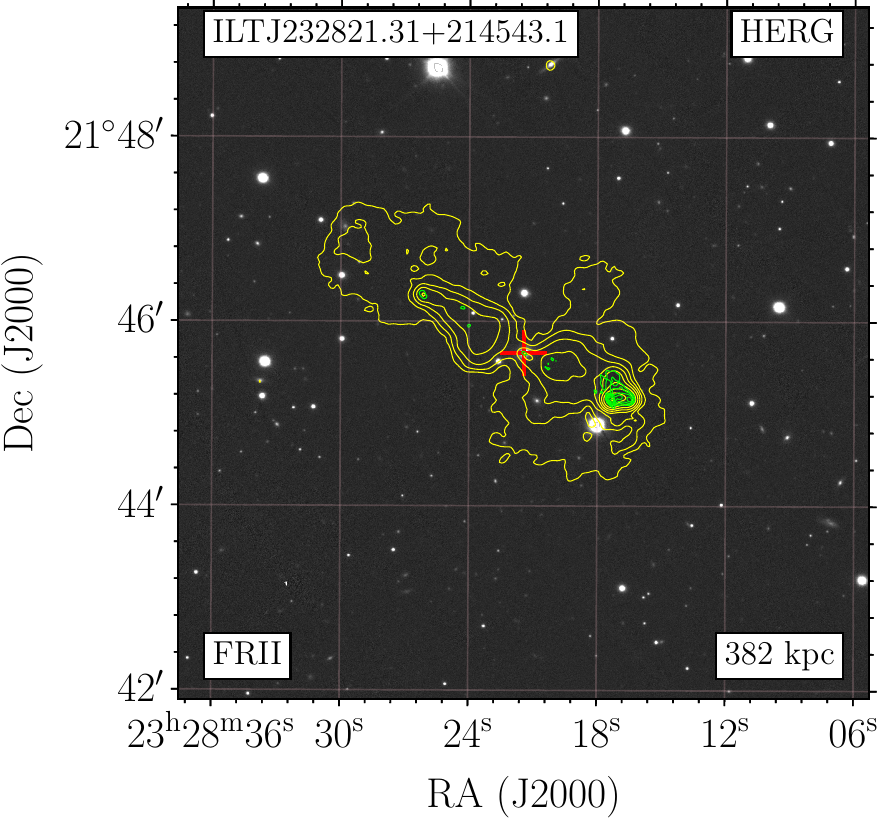}
    \includegraphics[width=0.65\columnwidth]{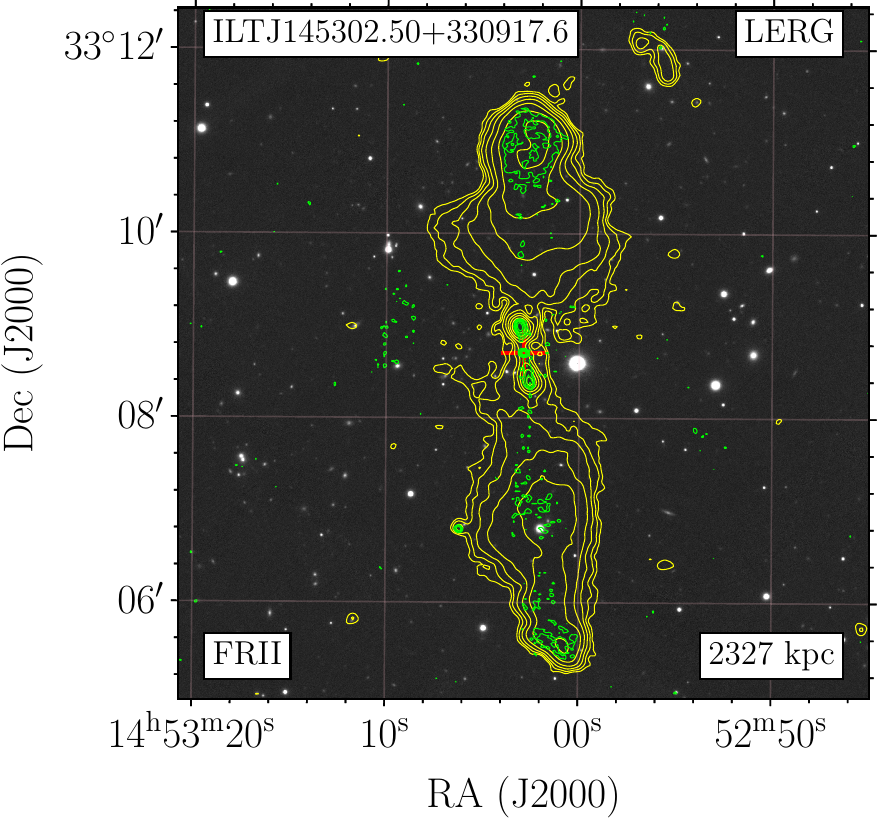}
    \caption{Continued from Fig. \ref{fig:images}. The bottom panel shows examples of restarted candidate sources, while the penultimate panel displays remnant candidates. Contours increase logarithmically by factors of 2, starting from 3$\sigma$.}
    \label{fig:images2}
\end{figure*}

\bsp	
\label{lastpage}
\end{document}